\documentclass[pra, aps, reprint, longbibliography] {revtex4-1}
\usepackage{times, mathrsfs, amsmath, amsfonts, graphics, graphicx, cancel, color, theorem, bbm, mathtools, amssymb}

\usepackage[english]{babel}
\usepackage[margin=.75in]{geometry}
\usepackage[T1]{fontenc}
\usepackage{xfrac,relsize,float}
\usepackage[unicode=true, breaklinks=false, pdfborder={0 0 1}, backref=false, colorlinks=true, linkcolor=blue, citecolor=blue]{hyperref}
\usepackage{physics}

\definecolor{Blue}{rgb}{0,0,1}
\definecolor{Red}{rgb}{1,0,0}
\definecolor{Green}{rgb}{0,1,0}
\definecolor{darkgreen}{rgb}{0,.7,0}
\definecolor{Purp}{rgb}{.2,0,.2}
\definecolor{white}{rgb}{1,1,1}

\begin{document}
\title{Exploiting the Causal Tensor Network Structure of Quantum Processes \\ to Efficiently Simulate Non-Markovian Path Integrals}
\author{Mathias R. J{\o}rgensen${}^{1}$}
    \email{matrj@fysik.dtu.dk}
\author{and Felix A. Pollock${}^{2}$}
    \email{felix.pollock@monash.edu}
\affiliation{${}^{1}$ Department of Physics, Technical University of Denmark, 2800 Kongens Lyngby, Denmark}
\affiliation{${}^{2}$ School of Physics and Astronomy, Monash University, Clayton, Victoria 3800, Australia}

\date{\today}
\begin{abstract}
In the path integral formulation of the evolution of an open quantum system coupled to a Gaussian, non-interacting environment, the dynamical contribution of the latter is encoded in an object called the influence functional.
Here, we relate the influence functional to the process tensor -- a more general representation of a quantum stochastic process -- describing the evolution.
We then use this connection to motivate a tensor network algorithm for the simulation of multi-time correlations in open systems, building on recent work where the influence functional is represented in terms of time evolving matrix product operators.
By exploiting the symmetries of the influence functional, we are able to use our algorithm to achieve orders-of-magnitude improvement in the efficiency of the resulting numerical simulation.
Our improved algorithm is then applied to compute exact phonon emission spectra for the spin-boson model with strong coupling,
demonstrating a significant divergence from spectra derived under commonly used assumptions of memorylessness.
\end{abstract}
\maketitle

\textit{Introduction}. --
All nanoscale quantum systems are open, meaning they inevitably interact with their environments,
exchanging energy and generating correlations.
If the system and its environment remain approximately uncorrelated,
then the reduced system dynamics is well described by a Markovian model~\cite{BreuerPetruccione2002,Weiss2012,Carmichael2003}.
However, in physical systems such as photosynthetic complexes, nanoscale lasers and quantum thermal machines~\cite{Mujica-Martinez2013PRE,McCutcheon2016PRA,Newmann2017PRE},
the need to go beyond a Markovian description has long been recognized,
and techniques accounting for non-Markovian physics have been developed,
with greater or lesser breadth of applicability.
Analytical methods involving time-local equations of motion exist,
but tend to be highly restricted to specific parameter regimes~\cite{Vacchini2010PRA,Lee2012JCP,Fruchtman2016NSR}.
Exact simulation often requires numerical methods,
e.g. discrete path integrals~\cite{Makri1995JCP-1,Makri1995JCP-2,Nalbach2011NJP,Dattani2012AIP,Strathearn2017NJP},
time non-local memory kernels~\cite{Geva2003JCP,Cohen2011PRB,Cerrillo2014PRL,Rosenbach2016NJP,Buser2017PRA,Gelzinis2017JCP,Pollock2018Qtm},
hierarchical equations of motion~\cite{Tanimura2006JPSJ,Strumpfer2012JCTC}
and others~\cite{Bulla2008RMP,Chen2017JCP-1,Chen2017JCP-2,Cerrillo2014PRL}.
Overall these methods tend to scale unfavourably with both the simulation time and the system size~\cite{deVega2017RMP},
making them inapplicable to important processes involving large complexes or when long time dynamics is important.

Recently, tensor network methods have been applied to the simulation~\cite{Chin2010JMP,Prior2010PRL,SchroederPRB2016,WallPRA2016}
and characterization~\cite{Pollock2018PRA, Luchnikovarx2018} of open quantum dynamics. 
Physically, these methods incorporate the fact that typical open quantum systems are only finitely correlated with their environments,
massively reducing their description~\cite{OrusAnP2014}.
In particular, Strathearn et al.~\cite{Strathearn2018NC} reformulated the discrete path integral for open systems
with Gaussian environments in terms of matrix product operators;
the resulting time-evolving matrix product operator (TEMPO) algorithm is numerically exact and has an efficiency comparable to other state of the art methods.
By effectively only considering the most important non-Markovian contributions to the dynamics,
the algorithm circumvents the exponential memory scaling of the bare path integral representation,
in a similar spirit to earlier path-filtering techniques~\cite{SimCPC1997, SimJCP2001, DattaniCPC2013}.
Motivated by this success, it is natural to ask if tensor network methods can be efficiently generalized from the simulation of reduced system density operators, to the simulation of general non-Markovian processes and multi-time correlations, which typically require many realizations of the dynamics to characterize.

In this Letter, we propose such a generalization, by making a formal connection between the path integral structure and the recently developed process tensor framework for characterizing general non-Markovian quantum processes~\cite{Pollock2018PRA}.
We then use this to argue for an alternative formulation of the TEMPO algorithm,
where we exploit the symmetry of the underlying tensor network to better account for the causal structure inherent in the dynamics. This not only allows us to efficiently compute multi-time correlation functions -- the simulation need only run once to extract all multi-time observable properties -- but also opens the door to the simulation of more general models.
Our alternative formulation is demonstrated to significantly improve the efficiency of the method,
which we use to straightforwardly compute non-Markovian emission spectra
for the spin-boson model, beyond the point where the commonly used quantum regression theorem breaks down~\cite{LiPhR2018}.

\begin{figure}
	\includegraphics[width=7.5cm]{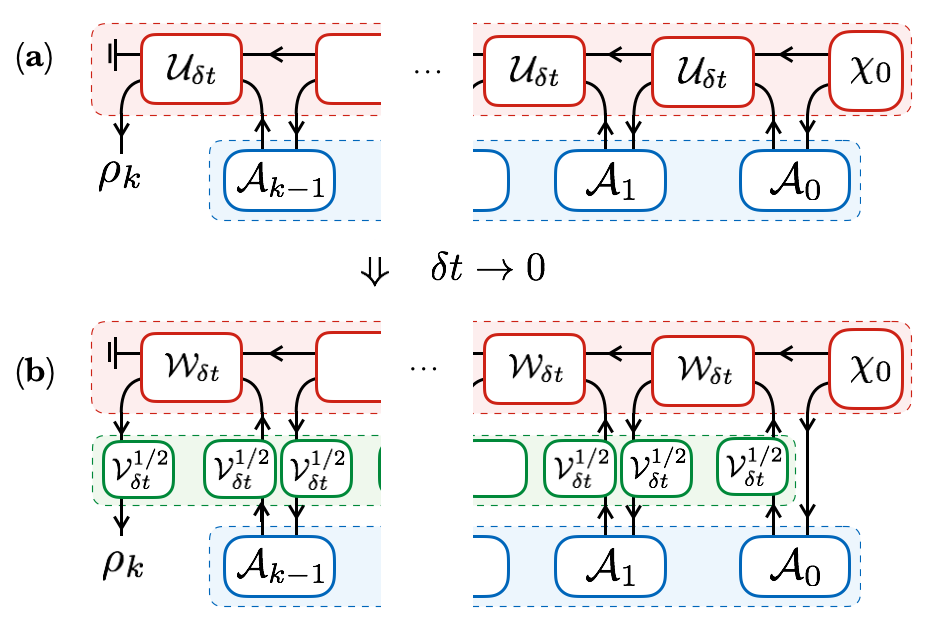}
	\caption{(Color online) (a) An arbitrary process with interventions can be represented as a matrix product form tensor network, the process tensor (upper row), that contracts with a filter function consisting of a sequence of superoperators (lower row).
	This makes it possible to separate implemented control operations from the underlying uncontrolled process.
	(b) In the infinitesimal time step limit,
	the uncontrolled process can be further decomposed into free evolution of the system (middle row) and a generalized
	influence functional capturing the influence of the environment.}
	\label{fig:StateFunction}
\end{figure}

\textit{Process tensor framework.} -- 
We consider stationary unitary dynamics of the system $S$ we are interested in along with its environment $E$,
and suppose the system is transformed by superoperators $\mathcal{A}_j$
at the discrete times $\left\lbrace t_{k-1}, ...,t_{0}\right\rbrace$,
which we take to be at evenly spaced intervals of $\delta t = t_j - t_{j-1}$.
In an experiment, these superoperators could correspond to actual interventions on the system as it evolves, i.e. unitary rotations, measurements etc., in which case they are completely-positive and, if the interventions are not conditional on a particular measurement outcome, trace preserving. Otherwise, the set $\{\mathcal{A}_j\}$ could represent more abstract transformations useful in the computation of physical quantities
such as operator expectation values or emission spectra.
The reduced, and potentially subnormalized, state of the system at time $t_{k}$ is given by
\begin{equation} \label{eq:QuantumProcess}
    \begin{aligned}
        \rho_{k}(\{\mathcal{A}_j\}) \ = \ 
        \tr_{E} \left\lbrace \mathcal{U}_{\delta t}\mathcal{A}_{k-1} \ ...
        \ \mathcal{U}_{\delta t} \mathcal{A}_{0} \left[ \chi_{0} \right] \right\rbrace \ ,
    \end{aligned}
\end{equation}
where $\mathcal{U}_{\delta t}$ is a superoperator representation of the unitary evolution of duration $\delta t$, i.e. $\mathcal{U}_{\delta t}[\rho] = U_{\delta t} \rho U_{\delta t}^\dagger$, with $U_{\delta t}$ a unitary matrix,
and $\chi_{0}$ is the initial system-environment state.
The inclusion of intermediate transformations makes it possible for us to consider a much broader class of physical properties than free evolution of the density operator
(corresponding to $\mathcal{A}_j = \mathcal{I}$ the identity superoperator~$\forall j$) would allow.

Since the state at time $t_k$ in Eq.~\eqref{eq:QuantumProcess} is linearly related to each of the set of superoperators $\{\mathcal{A}_j\}$ it can be written as a linear function of the tensor product of their Choi state representations $\textbf{A}_{k-1:0} = \mathsf{A}_{k-1} \otimes \dots \otimes \mathsf{A}_{1} \otimes \mathsf{A}_{0}$, with $\mathsf{A}_j:= \sum_{sr}\mathcal{A}_j[\ketbra{s}{r}]\otimes\ketbra{s}{r}$ obtained via the Choi-Jamio{\l}kowski isomorphism~\cite{MilzOSD2017,ChuangNielsen2011,Wilde2017}; here, $\{\ket{s}\}$ forms an orthonormal basis for $S$. Specifically, $\rho_{k}(\{\mathcal{A}_j\}) =  {\rm tr}_{k-1:0}\{\Upsilon_{k:0}(\mathbbm{1}_k\otimes\textbf{A}_{k-1:0}^T)\}$, with the trace over all subsystems on which $\textbf{A}_{k-1:0}$ acts. As we detail explicitly in Appendix~\ref{app:proc}, 
\begin{align}\label{eq:processtensor}
	    \Upsilon_{k:0}  =&\!\!\!\!   \sum_{\vec{s}',\vec{r}',\vec{s},\vec{r}}\!\!\!\tr \left\lbrace  \mathcal{U}_{\delta t}^{(s'_{k}, r'_{k},s_{k-1}, r_{k-1})} \!\!\!\!\!\! \dots \ \mathcal{U}_{\delta t}^{(s'_1, r'_1,s_0, r_0)} \!\left[ \chi^{(r'_0,s'_0)}_0 \right]  \right\rbrace \nonumber\\
	    & \qquad\times \ketbra{s'_k s_{k-1}\dots s'_1 s_0 s'_0}{r'_k r_{k-1}  \dots r'_1 r_0 r'_0}  ,
\end{align}
with environment superoperators $\mathcal{U}_{\delta t}^{(s', r',s, r)}[\rho^E] = \bra{s'}U_{\delta t} (\ketbra{s}{r} \otimes \rho^E) U^\dagger_{\delta t} \ket{r'}$ and operators $\chi^{(r',s')}_0 = \bra{r'}\chi_0\ket{s'}$, is the Choi representation of the \emph{process tensor}~\cite{Pollock2018PRA}, a many-body operator (on $2k+1$ copies of $S$) containing all information about the system's evolution that is independent of the transformations $\{\mathcal{A}_j\}$. Correlations between subsystems of $\Upsilon_{k:0}$ correspond to temporal correlations between observables, and a representation in terms of process tensors has been shown to consistently generalize stochastic processes, and related notions such as the Markov property and Markov order, to the quantum case~\cite{Milzarx2017, Pollock2018PRL, Taranto2019PRL, Taranto2019PRA}. The process tensor is illustrated graphically in Fig.~\ref{fig:StateFunction},
and can be thought of simply as a sequence of correlated maps on the system~\cite{SakuldeeJPA2018}.
Unlike in a conventional open quantum systems picture, where density operators are mapped to density operators,
this operational formulation stresses that the proper input to a quantum process is the set of interventions $\textbf{A}_{k-1:0}$,
and that the intermediate dynamics, and the initial state, are features of the process itself.

\textit{Gaussian influence functional.} --
Here, we consider the specific structure of the process tensor for systems interacting with Gaussian environments,
where the system-environment Hamiltonian and initial state
depend at most quadratically on environment creation and annihilation operators.
For concreteness, we focus on spin-boson type models,
but our results would extend to fermionic baths as well~\cite{AtlandSimon2010}.
Working in natural units ($\hbar=k_{B}=1$),
we consider a spin system, with Hilbert space dimension $d$,
interacting linearly with a bath of harmonic oscillators described by the Hamiltonian
$H = H_{0} + H_{B}$.
Here, $H_{0}$ describes the free spin system and the full bath influence is collected in
$H_{B} = \hat{s} \sum_{n} \left( g_{n}\hat{a}_{n} + g_{n}^{*}\hat{a}_{n}^{\dagger} \right)+  \sum_{n} \omega_{n} \hat{a}_{n}^{\dagger}\hat{a}_{n}$.
A bath mode $n$ has energy $\omega_{n}$,
and is created (annihilated) by the bosonic operator $\hat{a}_{n}^{\dagger}$ ($\hat{a}_{n}$).
The system operator $\hat{s}$ interacts with the bath with coupling constants $g_{n}$.
Additional linear interaction terms to different system operators could be included, as long as all these system operators commute. For simplicity, we take the initial state to be product, such that $\chi_{0} = \rho_{0} \otimes \tau_{\beta}$,
with the environment initially described by a thermal state 
$\tau_{\beta} = \exp \left[-\beta \sum \omega_n a^\dagger_n a_n \right]/ \mathcal{Z}$ at inverse temperature $\beta$,
where $\mathcal{Z} = \tr\{\exp \left[-\beta \sum \omega_n a^\dagger_n a_n \right]\}$.
This choice is not essential,
and other, possibly correlated, Gaussian initial states of the environment could be considered.

In the limit that the time difference $\delta t$ is small, the generated unitary dynamics
can be approximately separated into contributions arising from $H_0$ and $H_B$ as $\mathcal{U}_{\delta t} \simeq \mathcal{V}^{1/2}_{\delta t} \mathcal{W}_{\delta t}\mathcal{V}^{1/2}_{\delta t}$, where $\mathcal{V}_{\delta t}$ describes the free dynamics of $S$ and $\mathcal{W}_{\delta t}$ describes the environment influence.
The discrepancy between the approximate unitary maps and the actual ones vanishes as $\mathcal{O}(\delta t^{3})$ for this symmetric decomposition~\cite{Trotter1959PAMS}.
Note that the approximate model does not break unitarity,
and so corresponds to a valid physical process independently of step size.
Moreover, since the Hamiltonian only contains a single interaction term,
the interaction unitary preserves the eigenbasis of the corresponding system operator $\hat{s} = \sum_{s}\lambda_s \ketbra{s}{s}$: $\bra{s'}\mathcal{W}_{\delta t}[\ketbra{s}{r}]\ket{r'} = \delta_{ss'}\delta_{rr'}\mathcal{W}_{\delta t}^{(s, r)}$.
Together with the decomposition of unitary maps, this allows us to write the approximate process tensor Choi state as
\begin{equation} \label{eq:approxChoistate}
	\tilde{\Upsilon}_{k:0} \ = \ \left(\mathcal{V}^{1/2}_{\delta t} \otimes {\mathcal{V}^*_{\delta t} }^{1/2}\right)^{\otimes k}
	\left[ \mathcal{F}_{k:0} \right]  \otimes \rho_{0} \ ,
\end{equation}
where $\mathcal{F}_{k:0}$ is an operator representation of the discretized Feynman-Vernon influence functional \cite{FeynmanAnP1963}
encoding environment induced correlations
\begin{equation}\label{eq:infuncstate}
    \begin{aligned}
	    \mathcal{F}_{k:0}  =&   \sum_{\vec{s},\vec{r}}\tr_{E} \left\lbrace  \mathcal{W}_{\delta t}^{(s_{k}, r_{k})}  \!\dots  \mathcal{W}_{\delta t}^{(s_1, r_1)} \left[ \tau_{\beta} \right]  \right\rbrace\\
	    & \qquad\times \ketbra{s_k s_k \dots s_1 s_1}{r_k r_k \dots r_1 r_1} \ .
    \end{aligned}
\end{equation}

For Gaussian environments, the bath degrees of freedom can be traced over analytically using standard path integral techniques~\cite{Makri1995JCP-1, Makri1995JCP-2, DiosiPRL2014, Strathearn2017NJP}.
In this case, introducing the $d^{2}$ compound indices $\alpha=(s,r)$,
an element of the influence functional $\mathcal{F}_{k:0}^{\alpha_{k}...\alpha_{1}}:= \bra{s_k s_k \dots s_1 s_1}\mathcal{F}_{k:0}\ket{r_k r_k \dots r_1 r_1}$ can be decomposed as
\begin{equation}\label{eq:influencefunc}
	\mathcal{F}_{k:0}^{\alpha_{k}...\alpha_{1}} \ = \ 
     \prod_{i=1}^{k} \prod_{j=1}^{i} \left[b_{(i-j)}\right]^{\alpha_{i}\alpha_{j}} \ ,
\end{equation}
where $b_{(i-j)}$ is called an influence tensor; the exact form, which can often be approximated by an analytic function~\cite{DattaniQPL2012}, is given in Appendices~\ref{app:inffunc}~and~\ref{app:spinboson} along with a full derivation of Eqs.~\eqref{eq:infuncstate}~and~\eqref{eq:influencefunc}.
The influence tensors connect the dynamics around time step $i$ with that around step $j$,
quantifying the temporal correlations mediated by the environment between those two points;
that is, they describe memory effects.
Since the Hamiltonian is time-independent, the individual tensors $\left[b_{l}\right]$ depend only on the temporal separation
$l \delta t$, simplifying the potential complexity considerably. However, since the influence functional is a $k$ index tensor, it is still potentially exponentially complex; we now show how viewing Eq.~\eqref{eq:influencefunc} as a tensor network can make its calculation more tractable.

\begin{figure}
	\includegraphics[width=8.0cm]{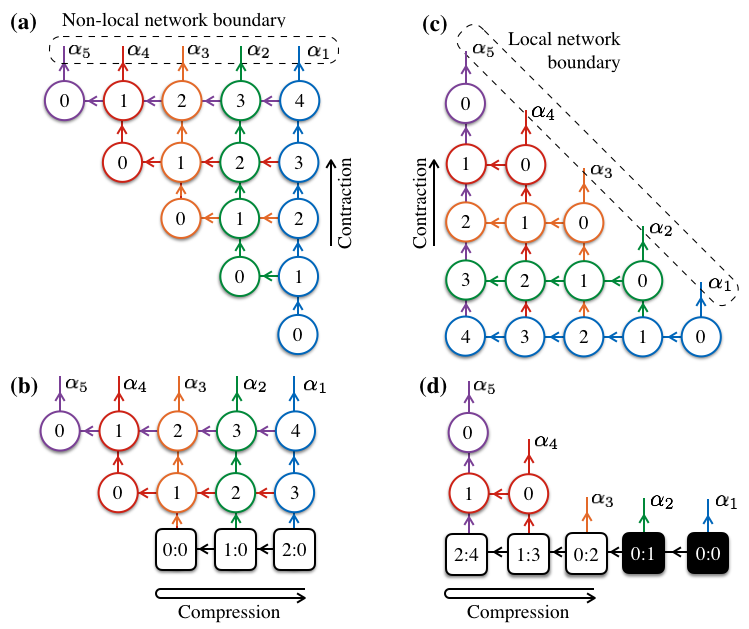}
	\caption{(Color online)
	Tensor network representation of the influence functional on five time steps,
	with nodes representing influence tensors and labelled by time step separation.
	Before contraction, indices are constrained to be equal along rows and columns in the network;
	the open boundary can therefore be shifted to any tensor in the same column (panels (a) and (b)) or row (panels (c) and (d)).
	(a) With the non-local boundary choice of Ref.~\cite{Strathearn2018NC},
	the free indices are attached to influence tensors encoding memory effects over all different time-scales.
	(b) The full network is contracted iteratively from below, row by row, down to a boundary MPO,
	with the full influence functional changing at each step.  
	(c) With the local boundary choice, the free indices are always attached to the time-local influence tensors.
	(d) Contraction proceeds as indicated, with the causal influence of each open leg sequentially incorporated into the wider network.
	The influence functional on an open leg is fixed once the corresponding layer has been contracted over.}
	\label{fig:network}
\end{figure}

\textit{Tensor network simulation.} --
In many cases, the environment interaction produces only finite length correlations in $\mathcal{F}_{k:0}$, a fact used by the authors of Ref.~\cite{Strathearn2018NC} to circumvent the exponential complexity growth by representing it efficiently in terms of
matrix product operators (MPOs)~\cite{Schollwock2011AP}.
To introduce this representation we first extend our two-index influence tensors into three-index tensors as
$\left[b_{(i-j)}\right]^{\gamma \alpha_{i}}{}_{\alpha_{j}} 
:= \delta^{\gamma}_{\alpha_{j}} \left[b_{(i-j)}\right]^{\alpha_{i}}{}_{\alpha_{j}}$,
where by convention an upper and a lower repeated index in a product of tensors is summed over
(otherwise, tensor elements differing only through raising or lowering are treated as equal).
In terms of these, we then define the \textit{non-local} time-evolving MPOs
\begin{equation}\label{eq:IFnonlocal}
    \begin{aligned}
        \mathcal{F}_{k:0}^{\alpha_{k} ... \alpha_{1}}
        & =
        \prod_{i=1}^{k}
        \left[b_{0} \right]^{\alpha_{i}}_{\beta_{1}}
        \prod_{j=1}^{i-2} \left[b_{j} \right]^{\beta_{j}}{}^{\alpha_{i-j}}_{\beta_{j+1}}
        \left[b_{i-1} \right]^{\beta_{i-1}\alpha_{1}}  \ \ \ \ \ \  \\
        & =
        \prod_{i=1}^{k} \ \
        \begin{minipage}[h]{0.0\linewidth}
	    \vspace{0pt}
        \includegraphics[width=4cm]{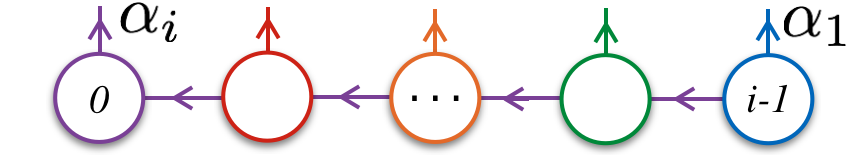}
        \end{minipage}\qquad\qquad\qquad\qquad\qquad\quad,
    \end{aligned}
\end{equation}
where the outgoing (ingoing) arrows in the graphical representation indicate upper (lower) indices,
and lines running through a given row or column are fixed to have the same index through Kronecker deltas.
At the right boundary of the tensors shown in the second line of Eq.~\eqref{eq:IFnonlocal},
we end up with a redundant lower index which we can trace over,
while at the left boundary we impose that the two upper indices be equal.
The full influence functional can then be constructed by iteratively multiplying such MPOs.
If we label the individual MPOs in the product by $G^{\alpha_{i}...\alpha_{1}}$,
then we can express the iterative multiplication as
\begin{equation}\label{eq:nonlocalprop}
    \begin{aligned}
        \mathcal{F}_{k:0}^{\alpha_{k}...\alpha_{1}}
        \ = \ \mathcal{G}^{\alpha_{k} \alpha_{k-1}...\alpha_{1}}_{\ \ \ \ \beta_{k-1}...\beta_{1}}
        \mathcal{F}_{k-1:0}^{\beta_{k-1}...\beta_{1}},
    \end{aligned}
\end{equation}
with $\mathcal{G}^{\alpha_{k} \alpha_{k-1}...\alpha_{1}}_{\ \ \ \ \beta_{k-1}...\beta_{1}}
    := G^{\alpha_{k}...\alpha_{1}} \delta^{\alpha_{k-1}}_{\beta_{k-1}}... \delta^{\alpha_{1}}_{\beta_{1}}$;
this is represented graphically by the two-dimensional tensor network shown in Fig.~\ref{fig:network}a.
Conceptually the use of time-evolving MPOs,
allows the state of the system to be propagated by updating indices to encode memory effects from the past process.
This type of propagation is analogous to a description in terms of a time non-local memory kernel, since the open legs are connected to tensors describing the influence of the state at various points in its history~\footnote{This correspondence is not precise, however, and our usage of `(non-)local' should not be confused with that in the  context of memory kernel convolution.}.

A key insight of this paper is that the decomposition of the influence functional into MPOs is not unique.
Kronecker deltas implicit in Eqs.~\eqref{eq:IFnonlocal}~and~\eqref{eq:nonlocalprop} mean that the open leg in a given row or column in Fig.~\ref{fig:network} could be shifted to any tensor in that same row or column.
In particular, the causal structure of the process tensor motivates an alternative definition in terms of
\textit{local} time-evolving MPOs
\begin{equation} \label{eq:IFlocal}
    \begin{aligned}
        \mathcal{F}_{k:0}^{\alpha_{k} ... \alpha_{1}}
        & =
        \prod_{i=1}^{k} 
        \left[b_{k-i} \right]^{\alpha_{k}}_{\beta_{k-i}}
        \prod_{j=1}^{k-i-1} \left[b_{j} \right]^{\beta_{j+1}}{}^{\alpha_{j+1}}_{\beta_{j}}
        \left[b_{0} \right]^{\beta_{1} \alpha_{i}} \\
        & =
        \prod_{i=1}^{k} \ \
        \begin{minipage}[h]{0.0\linewidth}
	    \vspace{0pt}
        \includegraphics[width=4cm]{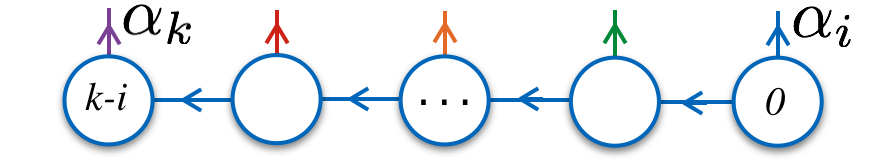}
        \end{minipage}\qquad\qquad\qquad\qquad\qquad\quad,
    \end{aligned}
\end{equation}
where now we end up with a redundant index at the left boundary, which we trace over,
and at the right boundary we impose the condition that the lower index must equal $\alpha_i$.
As with the non-local propagators,
the full influence functional is constructed iteratively by locally contracting MPOs.
Labelling the individual MPOs in the product by $C^{\alpha_{k}...\alpha_{i}}$,
the iterative multiplication can be expressed as (for $i \geq 1$)
\begin{equation}
    \begin{aligned}
        \tilde{\mathcal{F}}^{\alpha_{k}...\alpha_{1}}_{(k:i+1)}
        := \mathcal{C}^{\alpha_{k}...\alpha_{i+1}}_{\beta_{k}...\beta_{i+1}}
        \tilde{\mathcal{F}}^{\beta_{k}...\beta_{i+1} \alpha_{i}...\alpha_{1}}_{(k:i)},
    \end{aligned}
\end{equation}
with $\mathcal{C}^{\alpha_{k}...\alpha_{i}}_{\gamma_{k}...\gamma_{i}}
    := C^{\alpha_{k}...\alpha_{i}} \delta^{\alpha_{k}}_{\gamma_{k}}...\delta^{\alpha_{i}}_{\gamma_{i}}$
and $\tilde{\mathcal{F}}^{\alpha_{k}...\alpha_{1}}_{(k:1)}:=C^{\alpha_{k}...\alpha_{1}}$. The resulting network representation for the influence functional $\mathcal{F}_{k:0} = \tilde{\mathcal{F}}_{(k:k)}$ is shown in Fig.~\ref{fig:network}c.
Conceptually, the local time-evolving MPOs propagate the state by updating a set of effective memory space indices.
These indices describe how the environment is conditioned by the process at a given time, 
and this information on the conditioning can be propagated locally.
Since the process tensor, and hence the influence functional, has a well-defined causal structure, this conditioning only occurs from the past to the future. This means that, for a fixed evolution time, the size of the tensor to be updated decreases with each iteration.

In contracting the network, efficiency is achieved by incorporating a tensor compression procedure of the obtained boundary in each iteration.
In this work we make use of the singular value compression procedure (see Appendix~\ref{app:algorithm}) \cite{Schollwock2011AP,Strathearn2018NC}.
Roughly speaking, the local tensors of the boundary are subjected to a singular value decomposition.
The compression consists of discarding the eigenvalues below a specified singular value cutoff $\lambda_{c}$,
quantifying the hardness of the compression.
For the non-local algorithm (Fig. \ref{fig:network}a,b),
the tensors contracted in each iteration encode information about the influence of multiple time-steps on each other.
When correlations become smaller at longer time scales, as is typically the case, not all this information is relevant for describing the process as a whole.
The local algorithm (Fig. \ref{fig:network}c,d) incorporates this insight,
and separates out the most important contribution by including only the future influence of the environment at each timestep.
Generally the most local contributions have the largest singular values,
and therefore the separation means that the part of the boundary being propagated in the local case is less correlated,
which translates into a more efficient algorithm.

\begin{figure}
	\includegraphics[width=7.5cm]{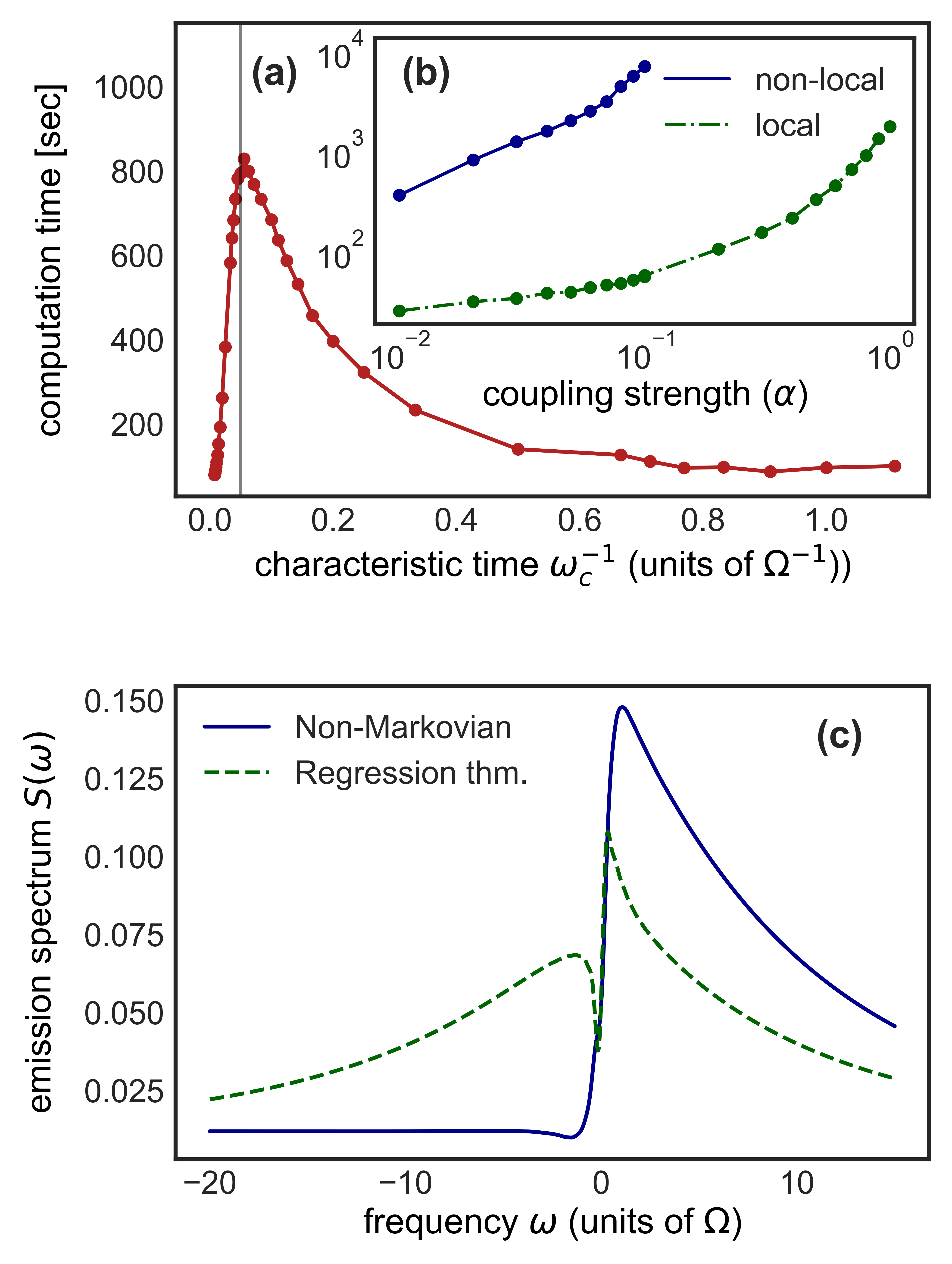}
	\caption{
	(a) Variation of computation time with the inverse of the cutoff frequency for the local algorithm at a coupling strength of $\alpha=0.7$,
	for an ohmic spectral density with $\omega_{c} = 10 \Omega$, $T = 0.01 \Omega$ and $\lambda_{c} = 10^{-6}$.
	(b) Comparison between the computation time of the non-local [Eq.~\eqref{eq:IFnonlocal}] and local [Eq.~\eqref{eq:IFlocal}]  time-evolving MPO algorithms for an Ohmic spectral density with $\omega_{c} = 10 \Omega$, $T = 0.01 \Omega$ and $\lambda_{c} = 10^{-6}$ as a function of coupling strength.
    (c) Steady state phonon emission spectrum at $\alpha=0.7$,
    the non-Markovian (numerically converged) spectrum is simulated using the local algorithm,
    and is compared with the spectrum obtained using the regression theorem (removing correlations across non-trivial superoperators).}
	\label{fig:complexity}
\end{figure}

\textit{Network complexity for a two level system.} --
We now turn to the specific simulation of the dynamics of a two-level system,
and compare the performance of the non-local and local algorithms.
Consider the free Hamiltonian $H_0 = \Omega \sigma_x/2$ and $\hat{s} = \sigma_z/2$,
where $\sigma_x$ and $\sigma_z$ are the usual Pauli operators.
The environment is fully characterized by its spectral density defined as
$J(\omega) = \sum_{n} |g_{n}|^{2} \delta\left( \omega - \omega_{n} \right)$~\cite{BreuerPetruccione2002}.
Here we consider a continuum bath model with the spectral density
$J(\omega) = (\alpha \omega_c/2) (\omega/\omega_{c})^{\nu} \exp \left( - \omega/\omega_{c} \right)$
with coupling strength $\alpha$, cutoff frequency $\omega_c$ and Ohmicity $\nu$, where for an Ohmic spectral density $\nu=1$.

The computational complexity is quantified by the computation time
required to contract a network of a certain size with a fixed singular value cutoff
(see Appendix~\ref{app:algorithm} for details on implementation). In general, this will depend on the overall magnitude of the influence functional, as well as the characteristic memory time quantifying how the elements of the influence tensors $b_{(i-j)}$ decrease in magnitude at large $|i-j|$. In Appendix~\ref{app:scaling}, we show that, for a fixed evolution time, the memory time goes as $\alpha/(\beta \omega_c)$ when $\omega_c$ is large, and that the overall coupling goes as $\alpha \omega_c t_{\rm max}^2/\beta$ when $\omega_c$ is small. 
In Fig.~\ref{fig:complexity}a,
we plot the computation time for the local and non-local algorithms as a function of coupling strength.
We find that the local representation outperforms the non-local one by one-to-two orders of magnitude, even at weak coupling,
and that the improvement increases at larger coupling strengths (in Appendix~\ref{app:efficiency} we show that there is an advantage everywhere across a wide range of parameters).
Furthermore, in Fig.~\ref{fig:complexity}b, we illustrate that this computational efficiency is maintained as the characteristic timescale of the bath (the inverse of the environment cutoff frequency) is varied, consistent with our predictions.

It should be kept in mind that,
unlike in most other quantum simulation methods, including the original TEMPO algorithm,
the object we are computing is the full process tensor,
which encodes all multi-time correlations, and from which a host of properties can be extracted efficiently.
In particular, we can compute the steady state emission spectrum
$S(\Delta \omega) =  \text{Re} \left[\int_{0}^{\infty} d\tau ( g^{(1)}(\tau)-g^{(1)}(\infty)) e^{-i \Delta \omega \tau} \right]$,
defined in terms of the two-point correlation function 
$g^{(1)}(\tau) = \lim_{t\rightarrow \infty} \left\langle \sigma^{\dagger}(t+\tau)\sigma(t) \right\rangle$.
The two-point correlation function is defined in terms of the raising and lowering operators on the spin system.
To compute it, we take all superoperators which the process tensor acts on to be the identity superoperator
$\mathcal{I}$ (with action $\mathcal{I}[\rho] = \rho$) except for two, which append a raising or lowering operator respectively.
In Fig.~\ref{fig:complexity}c, we study the physical effects of system-environment correlations by looking at the phonon emission spectrum.
We compare this with the spectrum computed using the quantum regression theorem, which approximates intermediate dynamics with that from an initial product state and is valid in the weak-coupling limit~\cite{McCutcheon2016PRA}.
The regression theorem correlations are obtained by breaking all correlations in the full process tensor Choi state across time steps at which the raising and lowering operators are evaluated.
In addition, we compare the exact spectrum with a fully Markovian process in which correlations are broken after each time step.
Fig.~\ref{fig:complexity}c shows that non-Markovian effects produce a phonon sideband in the spectrum at positive frequencies,
this is contrasted with the regression theorem result which gives a symmetric sideband structure.
Furthermore the fully Markovian spectrum contains no phonon sidebands, but rather a resonant emission peak.
These significant differences illustrates the importance of accounting for non-Markovian physics.

\textit{Conclusion.} -- In this Letter, we have established a direct connection between recent frameworks for characterizing general non-Markovian quantum processes and the path integral formulation of open quantum dynamics.
By relating the influence functional to a process tensor on an infinitesimal time grid, which has an explicit causal structure, we were able to build on recent progress in the classical simulation of Gaussian open quantum systems in terms of tensor networks.
Specifically, we showed that the speed of the TEMPO algorithm, when computing the multi-time properties encapsulated in the process tensor,
can be improved by orders-of-magnitude
by shifting the boundary of the corresponding tensor network from a temporally non-local to a local one.
Our contribution is immediately applicable to the
efficient simulation of realistic complex open systems, and additionally illustrates the utility of thinking about a time-local propagation of a conditioned environment space, rather than a description resembling the use of a non-local memory kernel.

The utility of the non-local TEMPO algorithm has been amply illustrated by computing the Ohmic localization transition,
and the dynamics of complex problems with multiple separated timescales \cite{Strathearn2018NC}.
The improved algorithm presented here is capable of exploring the same physics more efficiently,
and, in addition, easily extends to the computation of multi-time observables, of the sort crucial to describing, for example, multi-dimensional spectroscopy experiments~\cite{KassalSpectroscopy2014}. 
Moreover, relating path integral techniques to the more general process tensor formalism indicates how they might be generalized to more complex system-environment interactions, or even beyond the Gaussian regime.
Even within the spin-boson model, the freedom of boundary choice in Fig.~\ref{fig:network} that we have identified could be further exploited in other contexts. While the local choice appears optimal here, it may be that for other, structured spectral densities, where there are recurrent correlations, different boundary choices are more efficient, a point whose exploration we leave for future work.
\\

\begin{acknowledgments}
\textit{Acknowledgements.} --
FAP would like to thank A. Nazir, for asking one of the questions that motivated this work, and A. Strathearn, B. W. Lovett and P. Kirton for introducing and explaining the use of the TEMPO algorithm. MRJ was supported by the Independent Research Fund Denmark.
\end{acknowledgments}

\bibliography{bibliography}

\begin{thebibliography}{56}%
\makeatletter
\providecommand \@ifxundefined [1]{%
 \@ifx{#1\undefined}
}%
\providecommand \@ifnum [1]{%
 \ifnum #1\expandafter \@firstoftwo
 \else \expandafter \@secondoftwo
 \fi
}%
\providecommand \@ifx [1]{%
 \ifx #1\expandafter \@firstoftwo
 \else \expandafter \@secondoftwo
 \fi
}%
\providecommand \natexlab [1]{#1}%
\providecommand \enquote  [1]{``#1''}%
\providecommand \bibnamefont  [1]{#1}%
\providecommand \bibfnamefont [1]{#1}%
\providecommand \citenamefont [1]{#1}%
\providecommand \href@noop [0]{\@secondoftwo}%
\providecommand \href [0]{\begingroup \@sanitize@url \@href}%
\providecommand \@href[1]{\@@startlink{#1}\@@href}%
\providecommand \@@href[1]{\endgroup#1\@@endlink}%
\providecommand \@sanitize@url [0]{\catcode `\\12\catcode `\$12\catcode
  `\&12\catcode `\#12\catcode `\^12\catcode `\_12\catcode `\%12\relax}%
\providecommand \@@startlink[1]{}%
\providecommand \@@endlink[0]{}%
\providecommand \url  [0]{\begingroup\@sanitize@url \@url }%
\providecommand \@url [1]{\endgroup\@href {#1}{\urlprefix }}%
\providecommand \urlprefix  [0]{URL }%
\providecommand \Eprint [0]{\href }%
\providecommand \doibase [0]{http://dx.doi.org/}%
\providecommand \selectlanguage [0]{\@gobble}%
\providecommand \bibinfo  [0]{\@secondoftwo}%
\providecommand \bibfield  [0]{\@secondoftwo}%
\providecommand \translation [1]{[#1]}%
\providecommand \BibitemOpen [0]{}%
\providecommand \bibitemStop [0]{}%
\providecommand \bibitemNoStop [0]{.\EOS\space}%
\providecommand \EOS [0]{\spacefactor3000\relax}%
\providecommand \BibitemShut  [1]{\csname bibitem#1\endcsname}%
\let\auto@bib@innerbib\@empty
\bibitem [{\citenamefont {Breuer}\ and\ \citenamefont
  {Petruccione}(2002)}]{BreuerPetruccione2002}%
  \BibitemOpen
  \bibfield  {author} {\bibinfo {author} {\bibfnamefont {{H.-P.}}\ \bibnamefont
  {Breuer}}\ and\ \bibinfo {author} {\bibfnamefont {F.}~\bibnamefont
  {Petruccione}},\ }\href@noop {} {\emph {\bibinfo {title} {The Theory of Open
  Quantum Systems}}}\ (\bibinfo  {publisher} {Oxford University Press},\
  \bibinfo {year} {2002})\BibitemShut {NoStop}%
\bibitem [{\citenamefont {Weiss}(2012)}]{Weiss2012}%
  \BibitemOpen
  \bibfield  {author} {\bibinfo {author} {\bibfnamefont {U.}~\bibnamefont
  {Weiss}},\ }\href@noop {} {\emph {\bibinfo {title} {Quantum Dissipative
  Systems}}}\ (\bibinfo  {publisher} {World Scientific},\ \bibinfo {year}
  {2012})\BibitemShut {NoStop}%
\bibitem [{\citenamefont {Carmichael}(2003)}]{Carmichael2003}%
  \BibitemOpen
  \bibfield  {author} {\bibinfo {author} {\bibfnamefont {H.~J.}\ \bibnamefont
  {Carmichael}},\ }\href@noop {} {\emph {\bibinfo {title} {Statistical Methods
  in Quantum Optics 1: Master Equations and Fokker-Planck Equations}}}\
  (\bibinfo  {publisher} {Springer},\ \bibinfo {year} {2003})\BibitemShut
  {NoStop}%
\bibitem [{\citenamefont {Mujica-Martinez}\ \emph {et~al.}(2013)\citenamefont
  {Mujica-Martinez}, \citenamefont {Nalbach},\ and\ \citenamefont
  {Thorwart}}]{Mujica-Martinez2013PRE}%
  \BibitemOpen
  \bibfield  {author} {\bibinfo {author} {\bibfnamefont {C.~A.}\ \bibnamefont
  {Mujica-Martinez}}, \bibinfo {author} {\bibfnamefont {P.}~\bibnamefont
  {Nalbach}}, \ and\ \bibinfo {author} {\bibfnamefont {M.}~\bibnamefont
  {Thorwart}},\ }\bibfield  {title} {\enquote {\bibinfo {title} {Quantification
  of non-{Markovian} effects in the {Fenna-Matthews-Olson} complex},}\ }\href
  {\doibase 10.1103/PhysRevE.88.062719} {\bibfield  {journal} {\bibinfo
  {journal} {Phys. Rev. E}\ }\textbf {\bibinfo {volume} {88}},\ \bibinfo
  {pages} {062719} (\bibinfo {year} {2013})}\BibitemShut {NoStop}%
\bibitem [{\citenamefont {McCutcheon}(2016)}]{McCutcheon2016PRA}%
  \BibitemOpen
  \bibfield  {author} {\bibinfo {author} {\bibfnamefont {Dara P.~S.}\
  \bibnamefont {McCutcheon}},\ }\bibfield  {title} {\enquote {\bibinfo {title}
  {Optical signatures of non-{Markovian} behavior in open quantum systems},}\
  }\href {\doibase 10.1103/PhysRevA.93.022119} {\bibfield  {journal} {\bibinfo
  {journal} {Phys. Rev. A}\ }\textbf {\bibinfo {volume} {93}},\ \bibinfo
  {pages} {022119} (\bibinfo {year} {2016})}\BibitemShut {NoStop}%
\bibitem [{\citenamefont {Newman}\ \emph {et~al.}(2017)\citenamefont {Newman},
  \citenamefont {Mintert},\ and\ \citenamefont {Nazir}}]{Newmann2017PRE}%
  \BibitemOpen
  \bibfield  {author} {\bibinfo {author} {\bibfnamefont {David}\ \bibnamefont
  {Newman}}, \bibinfo {author} {\bibfnamefont {Florian}\ \bibnamefont
  {Mintert}}, \ and\ \bibinfo {author} {\bibfnamefont {Ahsan}\ \bibnamefont
  {Nazir}},\ }\bibfield  {title} {\enquote {\bibinfo {title} {Performance of a
  quantum heat engine at strong reservoir coupling},}\ }\href {\doibase
  10.1103/PhysRevE.95.032139} {\bibfield  {journal} {\bibinfo  {journal} {Phys.
  Rev. E}\ }\textbf {\bibinfo {volume} {95}},\ \bibinfo {pages} {032139}
  (\bibinfo {year} {2017})}\BibitemShut {NoStop}%
\bibitem [{\citenamefont {Vacchini}\ and\ \citenamefont
  {Breuer}(2010)}]{Vacchini2010PRA}%
  \BibitemOpen
  \bibfield  {author} {\bibinfo {author} {\bibfnamefont {Bassano}\ \bibnamefont
  {Vacchini}}\ and\ \bibinfo {author} {\bibfnamefont {Heinz-Peter}\
  \bibnamefont {Breuer}},\ }\bibfield  {title} {\enquote {\bibinfo {title}
  {Exact master equations for the non-{Markovian} decay of a qubit},}\ }\href
  {\doibase 10.1103/PhysRevA.81.042103} {\bibfield  {journal} {\bibinfo
  {journal} {Phys. Rev. A}\ }\textbf {\bibinfo {volume} {81}},\ \bibinfo
  {pages} {042103} (\bibinfo {year} {2010})}\BibitemShut {NoStop}%
\bibitem [{\citenamefont {Lee}\ \emph {et~al.}(2012)\citenamefont {Lee},
  \citenamefont {Moix},\ and\ \citenamefont {Cao}}]{Lee2012JCP}%
  \BibitemOpen
  \bibfield  {author} {\bibinfo {author} {\bibfnamefont {Chee~Kong}\
  \bibnamefont {Lee}}, \bibinfo {author} {\bibfnamefont {Jeremy}\ \bibnamefont
  {Moix}}, \ and\ \bibinfo {author} {\bibfnamefont {Jianshu}\ \bibnamefont
  {Cao}},\ }\bibfield  {title} {\enquote {\bibinfo {title} {Accuracy of second
  order perturbation theory in the polaron and variational polaron frames},}\
  }\href {\doibase 10.1063/1.4722336} {\bibfield  {journal} {\bibinfo
  {journal} {J. Chem. Phys.}\ }\textbf {\bibinfo {volume} {136}},\ \bibinfo
  {pages} {204120} (\bibinfo {year} {2012})}\BibitemShut {NoStop}%
\bibitem [{\citenamefont {Fruchtman}\ \emph {et~al.}(2016)\citenamefont
  {Fruchtman}, \citenamefont {Lambert},\ and\ \citenamefont
  {Gauger}}]{Fruchtman2016NSR}%
  \BibitemOpen
  \bibfield  {author} {\bibinfo {author} {\bibfnamefont {Amir}\ \bibnamefont
  {Fruchtman}}, \bibinfo {author} {\bibfnamefont {Neill}\ \bibnamefont
  {Lambert}}, \ and\ \bibinfo {author} {\bibfnamefont {Erik~M.}\ \bibnamefont
  {Gauger}},\ }\bibfield  {title} {\enquote {\bibinfo {title} {When do
  perturbative approaches accurately capture the dynamics of complex quantum
  systems?}}\ }\href {\doibase 10.1038/srep28204} {\bibfield  {journal}
  {\bibinfo  {journal} {Sci. Rep.}\ }\textbf {\bibinfo {volume} {6}},\ \bibinfo
  {pages} {28204} (\bibinfo {year} {2016})}\BibitemShut {NoStop}%
\bibitem [{\citenamefont {Makri}\ and\ \citenamefont
  {Makarov}(1995{\natexlab{a}})}]{Makri1995JCP-1}%
  \BibitemOpen
  \bibfield  {author} {\bibinfo {author} {\bibfnamefont {Nancy}\ \bibnamefont
  {Makri}}\ and\ \bibinfo {author} {\bibfnamefont {Dmitrii~E.}\ \bibnamefont
  {Makarov}},\ }\bibfield  {title} {\enquote {\bibinfo {title} {Tensor
  propagator for iterative quantum time evolution of reduced density matrices.
  {I. Theory}},}\ }\href {\doibase 10.1063/1.469508} {\bibfield  {journal}
  {\bibinfo  {journal} {J. Chem. Phys.}\ }\textbf {\bibinfo {volume} {102}},\
  \bibinfo {pages} {4600} (\bibinfo {year} {1995}{\natexlab{a}})}\BibitemShut
  {NoStop}%
\bibitem [{\citenamefont {Makri}\ and\ \citenamefont
  {Makarov}(1995{\natexlab{b}})}]{Makri1995JCP-2}%
  \BibitemOpen
  \bibfield  {author} {\bibinfo {author} {\bibfnamefont {Nancy}\ \bibnamefont
  {Makri}}\ and\ \bibinfo {author} {\bibfnamefont {Dmitrii~E.}\ \bibnamefont
  {Makarov}},\ }\bibfield  {title} {\enquote {\bibinfo {title} {Tensor
  propagator for iterative quantum time evolution of reduced density matrices.
  {II. Numerical methodology}},}\ }\href {\doibase 10.1063/1.469509} {\bibfield
   {journal} {\bibinfo  {journal} {J. Chem. Phys.}\ }\textbf {\bibinfo {volume}
  {102}},\ \bibinfo {pages} {4611} (\bibinfo {year}
  {1995}{\natexlab{b}})}\BibitemShut {NoStop}%
\bibitem [{\citenamefont {Nalbach}\ \emph {et~al.}(2011)\citenamefont
  {Nalbach}, \citenamefont {Ishizaki}, \citenamefont {Fleming},\ and\
  \citenamefont {Thorwart}}]{Nalbach2011NJP}%
  \BibitemOpen
  \bibfield  {author} {\bibinfo {author} {\bibfnamefont {Peter}\ \bibnamefont
  {Nalbach}}, \bibinfo {author} {\bibfnamefont {Akihito}\ \bibnamefont
  {Ishizaki}}, \bibinfo {author} {\bibfnamefont {Graham~R}\ \bibnamefont
  {Fleming}}, \ and\ \bibinfo {author} {\bibfnamefont {Michael}\ \bibnamefont
  {Thorwart}},\ }\bibfield  {title} {\enquote {\bibinfo {title} {Iterative
  path-integral algorithm versus cumulant time-nonlocal master equation
  approach for dissipative biomolecular exciton transport},}\ }\href {\doibase
  https://doi.org/10.1088/1367-2630/13/6/063040} {\bibfield  {journal}
  {\bibinfo  {journal} {New J. Phys.}\ }\textbf {\bibinfo {volume} {13}},\
  \bibinfo {pages} {063040} (\bibinfo {year} {2011})}\BibitemShut {NoStop}%
\bibitem [{\citenamefont {Dattani}(2012)}]{Dattani2012AIP}%
  \BibitemOpen
  \bibfield  {author} {\bibinfo {author} {\bibfnamefont {Nikesh~S.}\
  \bibnamefont {Dattani}},\ }\bibfield  {title} {\enquote {\bibinfo {title}
  {Numerical {Feynman} integrals with physically inspired interpolation: Faster
  convergence and significant reduction of computational cost},}\ }\href
  {\doibase 10.1063/1.3680607} {\bibfield  {journal} {\bibinfo  {journal} {AIP
  Adv.}\ }\textbf {\bibinfo {volume} {2}},\ \bibinfo {pages} {012121} (\bibinfo
  {year} {2012})}\BibitemShut {NoStop}%
\bibitem [{\citenamefont {Strathearn}\ \emph {et~al.}(2017)\citenamefont
  {Strathearn}, \citenamefont {Lovett},\ and\ \citenamefont
  {Kirton}}]{Strathearn2017NJP}%
  \BibitemOpen
  \bibfield  {author} {\bibinfo {author} {\bibfnamefont {A}~\bibnamefont
  {Strathearn}}, \bibinfo {author} {\bibfnamefont {B~W}\ \bibnamefont
  {Lovett}}, \ and\ \bibinfo {author} {\bibfnamefont {P}~\bibnamefont
  {Kirton}},\ }\bibfield  {title} {\enquote {\bibinfo {title} {Efficient
  real-time path integrals for non-{Markovian} spin-boson models},}\ }\href
  {\doibase 10.1088/1367-2630/aa8744} {\bibfield  {journal} {\bibinfo
  {journal} {New J. Phys.}\ }\textbf {\bibinfo {volume} {19}},\ \bibinfo
  {pages} {093009} (\bibinfo {year} {2017})}\BibitemShut {NoStop}%
\bibitem [{\citenamefont {Shi}\ and\ \citenamefont {Geva}(2003)}]{Geva2003JCP}%
  \BibitemOpen
  \bibfield  {author} {\bibinfo {author} {\bibfnamefont {Qiang}\ \bibnamefont
  {Shi}}\ and\ \bibinfo {author} {\bibfnamefont {Eitan}\ \bibnamefont {Geva}},\
  }\bibfield  {title} {\enquote {\bibinfo {title} {A new approach to
  calculating the memory kernel of the generalized quantum master equation for
  an arbitrary system-bath coupling},}\ }\href {\doibase 10.1063/1.1624830}
  {\bibfield  {journal} {\bibinfo  {journal} {J. Chem. Phys.}\ }\textbf
  {\bibinfo {volume} {119}},\ \bibinfo {pages} {12063--12076} (\bibinfo {year}
  {2003})}\BibitemShut {NoStop}%
\bibitem [{\citenamefont {Cohen}\ and\ \citenamefont
  {Rabani}(2011)}]{Cohen2011PRB}%
  \BibitemOpen
  \bibfield  {author} {\bibinfo {author} {\bibfnamefont {Guy}\ \bibnamefont
  {Cohen}}\ and\ \bibinfo {author} {\bibfnamefont {Eran}\ \bibnamefont
  {Rabani}},\ }\bibfield  {title} {\enquote {\bibinfo {title} {Memory effects
  in nonequilibrium quantum impurity models},}\ }\href {\doibase
  10.1103/PhysRevB.84.075150} {\bibfield  {journal} {\bibinfo  {journal} {Phys.
  Rev. B}\ }\textbf {\bibinfo {volume} {84}},\ \bibinfo {pages} {075150}
  (\bibinfo {year} {2011})}\BibitemShut {NoStop}%
\bibitem [{\citenamefont {Cerrillo}\ and\ \citenamefont
  {Cao}(2014)}]{Cerrillo2014PRL}%
  \BibitemOpen
  \bibfield  {author} {\bibinfo {author} {\bibfnamefont {Javier}\ \bibnamefont
  {Cerrillo}}\ and\ \bibinfo {author} {\bibfnamefont {Jianshu}\ \bibnamefont
  {Cao}},\ }\bibfield  {title} {\enquote {\bibinfo {title} {Non-{Markovian}
  dynamical maps: Numerical processing of open quantum trajectories},}\ }\href
  {\doibase 10.1103/PhysRevLett.112.110401} {\bibfield  {journal} {\bibinfo
  {journal} {Phys. Rev. Lett.}\ }\textbf {\bibinfo {volume} {112}},\ \bibinfo
  {pages} {110401} (\bibinfo {year} {2014})}\BibitemShut {NoStop}%
\bibitem [{\citenamefont {Rosenbach}\ \emph {et~al.}(2016)\citenamefont
  {Rosenbach}, \citenamefont {Cerrillo}, \citenamefont {Huelga}, \citenamefont
  {Cao},\ and\ \citenamefont {Plenio}}]{Rosenbach2016NJP}%
  \BibitemOpen
  \bibfield  {author} {\bibinfo {author} {\bibfnamefont {Robert}\ \bibnamefont
  {Rosenbach}}, \bibinfo {author} {\bibfnamefont {Javier}\ \bibnamefont
  {Cerrillo}}, \bibinfo {author} {\bibfnamefont {Susana~F}\ \bibnamefont
  {Huelga}}, \bibinfo {author} {\bibfnamefont {Jianshu}\ \bibnamefont {Cao}}, \
  and\ \bibinfo {author} {\bibfnamefont {Martin~B}\ \bibnamefont {Plenio}},\
  }\bibfield  {title} {\enquote {\bibinfo {title} {Efficient simulation of
  non-{Markovian} system-environment interaction},}\ }\href {\doibase
  10.1088/1367-2630/18/2/023035} {\bibfield  {journal} {\bibinfo  {journal}
  {New J. Phys.}\ }\textbf {\bibinfo {volume} {18}},\ \bibinfo {pages} {023035}
  (\bibinfo {year} {2016})}\BibitemShut {NoStop}%
\bibitem [{\citenamefont {Buser}\ \emph {et~al.}(2017)\citenamefont {Buser},
  \citenamefont {Cerrillo}, \citenamefont {Schaller},\ and\ \citenamefont
  {Cao}}]{Buser2017PRA}%
  \BibitemOpen
  \bibfield  {author} {\bibinfo {author} {\bibfnamefont {Maximilian}\
  \bibnamefont {Buser}}, \bibinfo {author} {\bibfnamefont {Javier}\
  \bibnamefont {Cerrillo}}, \bibinfo {author} {\bibfnamefont {Gernot}\
  \bibnamefont {Schaller}}, \ and\ \bibinfo {author} {\bibfnamefont {Jianshu}\
  \bibnamefont {Cao}},\ }\bibfield  {title} {\enquote {\bibinfo {title}
  {Initial system-environment correlations via the transfer-tensor method},}\
  }\href {\doibase 10.1103/PhysRevA.96.062122} {\bibfield  {journal} {\bibinfo
  {journal} {Phys. Rev. A}\ }\textbf {\bibinfo {volume} {96}},\ \bibinfo
  {pages} {062122} (\bibinfo {year} {2017})}\BibitemShut {NoStop}%
\bibitem [{\citenamefont {Gelzinis}\ \emph {et~al.}(2017)\citenamefont
  {Gelzinis}, \citenamefont {Rybakovas},\ and\ \citenamefont
  {Valkunas}}]{Gelzinis2017JCP}%
  \BibitemOpen
  \bibfield  {author} {\bibinfo {author} {\bibfnamefont {Andrius}\ \bibnamefont
  {Gelzinis}}, \bibinfo {author} {\bibfnamefont {Edvardas}\ \bibnamefont
  {Rybakovas}}, \ and\ \bibinfo {author} {\bibfnamefont {Leonas}\ \bibnamefont
  {Valkunas}},\ }\bibfield  {title} {\enquote {\bibinfo {title} {Applicability
  of transfer tensor method for open quantum system dynamics},}\ }\href
  {\doibase 10.1063/1.5009086} {\bibfield  {journal} {\bibinfo  {journal} {J.
  Chem. Phys.}\ }\textbf {\bibinfo {volume} {147}},\ \bibinfo {pages} {234108}
  (\bibinfo {year} {2017})}\BibitemShut {NoStop}%
\bibitem [{\citenamefont {Pollock}\ and\ \citenamefont
  {Modi}(2018)}]{Pollock2018Qtm}%
  \BibitemOpen
  \bibfield  {author} {\bibinfo {author} {\bibfnamefont {Felix~A.}\
  \bibnamefont {Pollock}}\ and\ \bibinfo {author} {\bibfnamefont {Kavan}\
  \bibnamefont {Modi}},\ }\bibfield  {title} {\enquote {\bibinfo {title}
  {Tomographically reconstructed master equations for any open quantum
  dynamics},}\ }\href {\doibase 10.22331/q-2018-07-11-76} {\bibfield  {journal}
  {\bibinfo  {journal} {{Quantum}}\ }\textbf {\bibinfo {volume} {2}},\ \bibinfo
  {pages} {76} (\bibinfo {year} {2018})}\BibitemShut {NoStop}%
\bibitem [{\citenamefont {Tanimura}(2006)}]{Tanimura2006JPSJ}%
  \BibitemOpen
  \bibfield  {author} {\bibinfo {author} {\bibfnamefont {Yoshitaka}\
  \bibnamefont {Tanimura}},\ }\bibfield  {title} {\enquote {\bibinfo {title}
  {Stochastic {Liouville}, {Langevin}, {Fokker-Planck}, and master equation
  approaches to quantum dissipative systems},}\ }\href {\doibase
  10.1143/JPSJ.75.082001} {\bibfield  {journal} {\bibinfo  {journal} {J. Phys.
  Soc. Jpn.}\ }\textbf {\bibinfo {volume} {75}},\ \bibinfo {pages} {082001}
  (\bibinfo {year} {2006})}\BibitemShut {NoStop}%
\bibitem [{\citenamefont {Str\"umpfer}\ and\ \citenamefont
  {Schulten}(2012)}]{Strumpfer2012JCTC}%
  \BibitemOpen
  \bibfield  {author} {\bibinfo {author} {\bibfnamefont {Johan}\ \bibnamefont
  {Str\"umpfer}}\ and\ \bibinfo {author} {\bibfnamefont {Klaus}\ \bibnamefont
  {Schulten}},\ }\bibfield  {title} {\enquote {\bibinfo {title} {Open quantum
  dynamics calculations with the hierarchy equations of motion on parallel
  computers},}\ }\href {\doibase 10.1021/ct3003833} {\bibfield  {journal}
  {\bibinfo  {journal} {J. Chem. Theory Comput.}\ }\textbf {\bibinfo {volume}
  {8}},\ \bibinfo {pages} {2808} (\bibinfo {year} {2012})}\BibitemShut
  {NoStop}%
\bibitem [{\citenamefont {Bulla}\ \emph {et~al.}(2008)\citenamefont {Bulla},
  \citenamefont {Costi},\ and\ \citenamefont {Pruschke}}]{Bulla2008RMP}%
  \BibitemOpen
  \bibfield  {author} {\bibinfo {author} {\bibfnamefont {Ralf}\ \bibnamefont
  {Bulla}}, \bibinfo {author} {\bibfnamefont {Theo~A.}\ \bibnamefont {Costi}},
  \ and\ \bibinfo {author} {\bibfnamefont {Thomas}\ \bibnamefont {Pruschke}},\
  }\bibfield  {title} {\enquote {\bibinfo {title} {Numerical renormalization
  group method for quantum impurity systems},}\ }\href {\doibase
  10.1103/RevModPhys.80.395} {\bibfield  {journal} {\bibinfo  {journal} {Rev.
  Mod. Phys.}\ }\textbf {\bibinfo {volume} {80}},\ \bibinfo {pages} {395}
  (\bibinfo {year} {2008})}\BibitemShut {NoStop}%
\bibitem [{\citenamefont {Chen}\ \emph
  {et~al.}(2017{\natexlab{a}})\citenamefont {Chen}, \citenamefont {Cohen},\
  and\ \citenamefont {Reichman}}]{Chen2017JCP-1}%
  \BibitemOpen
  \bibfield  {author} {\bibinfo {author} {\bibfnamefont {Hsing-Ta}\
  \bibnamefont {Chen}}, \bibinfo {author} {\bibfnamefont {Guy}\ \bibnamefont
  {Cohen}}, \ and\ \bibinfo {author} {\bibfnamefont {David~R.}\ \bibnamefont
  {Reichman}},\ }\bibfield  {title} {\enquote {\bibinfo {title} {Inchworm
  {Monte Carlo} for exact non-adiabatic dynamics. {I. Theory} and
  algorithms},}\ }\href {\doibase 10.1063/1.4974328} {\bibfield  {journal}
  {\bibinfo  {journal} {J. Chem. Phys.}\ }\textbf {\bibinfo {volume} {146}},\
  \bibinfo {pages} {054105} (\bibinfo {year} {2017}{\natexlab{a}})}\BibitemShut
  {NoStop}%
\bibitem [{\citenamefont {Chen}\ \emph
  {et~al.}(2017{\natexlab{b}})\citenamefont {Chen}, \citenamefont {Cohen},\
  and\ \citenamefont {Reichman}}]{Chen2017JCP-2}%
  \BibitemOpen
  \bibfield  {author} {\bibinfo {author} {\bibfnamefont {Hsing-Ta}\
  \bibnamefont {Chen}}, \bibinfo {author} {\bibfnamefont {Guy}\ \bibnamefont
  {Cohen}}, \ and\ \bibinfo {author} {\bibfnamefont {David~R.}\ \bibnamefont
  {Reichman}},\ }\bibfield  {title} {\enquote {\bibinfo {title} {Inchworm
  {Monte Carlo} for exact non-adiabatic dynamics. {II. Benchmarks} and
  comparison with established methods},}\ }\href {\doibase 10.1063/1.4974329}
  {\bibfield  {journal} {\bibinfo  {journal} {J. Chem. Phys.}\ }\textbf
  {\bibinfo {volume} {146}},\ \bibinfo {pages} {054106} (\bibinfo {year}
  {2017}{\natexlab{b}})}\BibitemShut {NoStop}%
\bibitem [{\citenamefont {de~Vega}\ and\ \citenamefont
  {Alonso}(2017)}]{deVega2017RMP}%
  \BibitemOpen
  \bibfield  {author} {\bibinfo {author} {\bibfnamefont {In\'es}\ \bibnamefont
  {de~Vega}}\ and\ \bibinfo {author} {\bibfnamefont {Daniel}\ \bibnamefont
  {Alonso}},\ }\bibfield  {title} {\enquote {\bibinfo {title} {Dynamics of
  non-{Markovian} open quantum systems},}\ }\href {\doibase
  10.1103/RevModPhys.89.015001} {\bibfield  {journal} {\bibinfo  {journal}
  {Rev. Mod. Phys.}\ }\textbf {\bibinfo {volume} {89}},\ \bibinfo {pages}
  {015001} (\bibinfo {year} {2017})}\BibitemShut {NoStop}%
\bibitem [{\citenamefont {Chin}\ \emph {et~al.}(2010)\citenamefont {Chin},
  \citenamefont {Rivas}, \citenamefont {Huelga},\ and\ \citenamefont
  {Plenio}}]{Chin2010JMP}%
  \BibitemOpen
  \bibfield  {author} {\bibinfo {author} {\bibfnamefont {Alex~W.}\ \bibnamefont
  {Chin}}, \bibinfo {author} {\bibfnamefont {{\'{A}}ngel}\ \bibnamefont
  {Rivas}}, \bibinfo {author} {\bibfnamefont {Susana~F.}\ \bibnamefont
  {Huelga}}, \ and\ \bibinfo {author} {\bibfnamefont {Martin~B.}\ \bibnamefont
  {Plenio}},\ }\bibfield  {title} {\enquote {\bibinfo {title} {Exact mapping
  between system-reservoir quantum models and semi-infinite discrete chains
  using orthogonal polynomials},}\ }\href {\doibase 10.1063/1.3490188}
  {\bibfield  {journal} {\bibinfo  {journal} {J. Math. Phys.}\ }\textbf
  {\bibinfo {volume} {51}},\ \bibinfo {pages} {092109} (\bibinfo {year}
  {2010})}\BibitemShut {NoStop}%
\bibitem [{\citenamefont {Prior}\ \emph {et~al.}(2010)\citenamefont {Prior},
  \citenamefont {Chin}, \citenamefont {Huelga},\ and\ \citenamefont
  {Plenio}}]{Prior2010PRL}%
  \BibitemOpen
  \bibfield  {author} {\bibinfo {author} {\bibfnamefont {Javier}\ \bibnamefont
  {Prior}}, \bibinfo {author} {\bibfnamefont {Alex~W.}\ \bibnamefont {Chin}},
  \bibinfo {author} {\bibfnamefont {Susana~F.}\ \bibnamefont {Huelga}}, \ and\
  \bibinfo {author} {\bibfnamefont {Martin~B.}\ \bibnamefont {Plenio}},\
  }\bibfield  {title} {\enquote {\bibinfo {title} {Efficient simulation of
  strong system-environment interactions},}\ }\href {\doibase
  10.1103/PhysRevLett.105.050404} {\bibfield  {journal} {\bibinfo  {journal}
  {Phys. Rev. Lett.}\ }\textbf {\bibinfo {volume} {105}},\ \bibinfo {pages}
  {050404} (\bibinfo {year} {2010})}\BibitemShut {NoStop}%
\bibitem [{\citenamefont {Schr\"oder}\ and\ \citenamefont
  {Chin}(2016)}]{SchroederPRB2016}%
  \BibitemOpen
  \bibfield  {author} {\bibinfo {author} {\bibfnamefont {Florian A. Y.~N.}\
  \bibnamefont {Schr\"oder}}\ and\ \bibinfo {author} {\bibfnamefont {Alex~W.}\
  \bibnamefont {Chin}},\ }\bibfield  {title} {\enquote {\bibinfo {title}
  {Simulating open quantum dynamics with time-dependent variational matrix
  product states: Towards microscopic correlation of environment dynamics and
  reduced system evolution},}\ }\href {\doibase 10.1103/PhysRevB.93.075105}
  {\bibfield  {journal} {\bibinfo  {journal} {Phys. Rev. B}\ }\textbf {\bibinfo
  {volume} {93}},\ \bibinfo {pages} {075105} (\bibinfo {year}
  {2016})}\BibitemShut {NoStop}%
\bibitem [{\citenamefont {Wall}\ \emph {et~al.}(2016)\citenamefont {Wall},
  \citenamefont {Safavi-Naini},\ and\ \citenamefont {Rey}}]{WallPRA2016}%
  \BibitemOpen
  \bibfield  {author} {\bibinfo {author} {\bibfnamefont {Michael~L.}\
  \bibnamefont {Wall}}, \bibinfo {author} {\bibfnamefont {Arghavan}\
  \bibnamefont {Safavi-Naini}}, \ and\ \bibinfo {author} {\bibfnamefont
  {Ana~Maria}\ \bibnamefont {Rey}},\ }\bibfield  {title} {\enquote {\bibinfo
  {title} {Simulating generic spin-boson models with matrix product states},}\
  }\href {\doibase 10.1103/PhysRevA.94.053637} {\bibfield  {journal} {\bibinfo
  {journal} {Phys. Rev. A}\ }\textbf {\bibinfo {volume} {94}},\ \bibinfo
  {pages} {053637} (\bibinfo {year} {2016})}\BibitemShut {NoStop}%
\bibitem [{\citenamefont {Pollock}\ \emph
  {et~al.}(2018{\natexlab{a}})\citenamefont {Pollock}, \citenamefont
  {Rodr\'{\i}guez-Rosario}, \citenamefont {Frauenheim}, \citenamefont
  {Paternostro},\ and\ \citenamefont {Modi}}]{Pollock2018PRA}%
  \BibitemOpen
  \bibfield  {author} {\bibinfo {author} {\bibfnamefont {Felix~A.}\
  \bibnamefont {Pollock}}, \bibinfo {author} {\bibfnamefont {C\'esar}\
  \bibnamefont {Rodr\'{\i}guez-Rosario}}, \bibinfo {author} {\bibfnamefont
  {Thomas}\ \bibnamefont {Frauenheim}}, \bibinfo {author} {\bibfnamefont
  {Mauro}\ \bibnamefont {Paternostro}}, \ and\ \bibinfo {author} {\bibfnamefont
  {Kavan}\ \bibnamefont {Modi}},\ }\bibfield  {title} {\enquote {\bibinfo
  {title} {Non-{Markovian} quantum processes: Complete framework and efficient
  characterization},}\ }\href {\doibase 10.1103/PhysRevA.97.012127} {\bibfield
  {journal} {\bibinfo  {journal} {Phys. Rev. A}\ }\textbf {\bibinfo {volume}
  {97}},\ \bibinfo {pages} {012127} (\bibinfo {year}
  {2018}{\natexlab{a}})}\BibitemShut {NoStop}%
\bibitem [{\citenamefont {Luchnikov}\ \emph {et~al.}(2019)\citenamefont
  {Luchnikov}, \citenamefont {Vintskevich}, \citenamefont {Ouerdane},\ and\
  \citenamefont {Filippov}}]{Luchnikovarx2018}%
  \BibitemOpen
  \bibfield  {author} {\bibinfo {author} {\bibfnamefont {I.~A.}\ \bibnamefont
  {Luchnikov}}, \bibinfo {author} {\bibfnamefont {S.~V.}\ \bibnamefont
  {Vintskevich}}, \bibinfo {author} {\bibfnamefont {H.}~\bibnamefont
  {Ouerdane}}, \ and\ \bibinfo {author} {\bibfnamefont {S.~N.}\ \bibnamefont
  {Filippov}},\ }\bibfield  {title} {\enquote {\bibinfo {title} {Simulation
  complexity of open quantum dynamics: Connection with tensor networks},}\
  }\href {\doibase 10.1103/PhysRevLett.122.160401} {\bibfield  {journal}
  {\bibinfo  {journal} {Phys. Rev. Lett.}\ }\textbf {\bibinfo {volume} {122}},\
  \bibinfo {pages} {160401} (\bibinfo {year} {2019})}\BibitemShut {NoStop}%
\bibitem [{\citenamefont {Or\'{u}s}(2014)}]{OrusAnP2014}%
  \BibitemOpen
  \bibfield  {author} {\bibinfo {author} {\bibfnamefont {Rom\'{a}n}\
  \bibnamefont {Or\'{u}s}},\ }\bibfield  {title} {\enquote {\bibinfo {title} {A
  practical introduction to tensor networks: Matrix product states and
  projected entangled pair states},}\ }\href {\doibase
  https://doi.org/10.1016/j.aop.2014.06.013} {\bibfield  {journal} {\bibinfo
  {journal} {Ann. Phys.}\ }\textbf {\bibinfo {volume} {349}},\ \bibinfo {pages}
  {117} (\bibinfo {year} {2014})}\BibitemShut {NoStop}%
\bibitem [{\citenamefont {{Strathearn}}\ \emph {et~al.}(2018)\citenamefont
  {{Strathearn}}, \citenamefont {{Kirton}}, \citenamefont {{Kilda}},
  \citenamefont {{Keeling}},\ and\ \citenamefont
  {{Lovett}}}]{Strathearn2018NC}%
  \BibitemOpen
  \bibfield  {author} {\bibinfo {author} {\bibfnamefont {A.}~\bibnamefont
  {{Strathearn}}}, \bibinfo {author} {\bibfnamefont {P.}~\bibnamefont
  {{Kirton}}}, \bibinfo {author} {\bibfnamefont {D.}~\bibnamefont {{Kilda}}},
  \bibinfo {author} {\bibfnamefont {J.}~\bibnamefont {{Keeling}}}, \ and\
  \bibinfo {author} {\bibfnamefont {B.~W.}\ \bibnamefont {{Lovett}}},\
  }\bibfield  {title} {\enquote {\bibinfo {title} {Efficient non-{Markovian}
  quantum dynamics using time-evolving matrix product operators},}\ }\href
  {\doibase 10.1038/s41467-018-05617-3} {\bibfield  {journal} {\bibinfo
  {journal} {Nat. Commun.}\ }\textbf {\bibinfo {volume} {9}},\ \bibinfo {pages}
  {3322} (\bibinfo {year} {2018})}\BibitemShut {NoStop}%
\bibitem [{\citenamefont {Sim}\ and\ \citenamefont {Makri}(1997)}]{SimCPC1997}%
  \BibitemOpen
  \bibfield  {author} {\bibinfo {author} {\bibfnamefont {Eunji}\ \bibnamefont
  {Sim}}\ and\ \bibinfo {author} {\bibfnamefont {Nancy}\ \bibnamefont
  {Makri}},\ }\bibfield  {title} {\enquote {\bibinfo {title} {Filtered
  propagator functional for iterative dynamics of quantum dissipative
  systems},}\ }\href {\doibase https://doi.org/10.1016/S0010-4655(96)00130-0}
  {\bibfield  {journal} {\bibinfo  {journal} {Comput. Phys. Commun.}\ }\textbf
  {\bibinfo {volume} {99}},\ \bibinfo {pages} {335} (\bibinfo {year}
  {1997})}\BibitemShut {NoStop}%
\bibitem [{\citenamefont {Sim}(2001)}]{SimJCP2001}%
  \BibitemOpen
  \bibfield  {author} {\bibinfo {author} {\bibfnamefont {Eunji}\ \bibnamefont
  {Sim}},\ }\bibfield  {title} {\enquote {\bibinfo {title} {Quantum dynamics
  for a system coupled to slow baths: On-the-fly filtered propagator method},}\
  }\href {\doibase 10.1063/1.1394208} {\bibfield  {journal} {\bibinfo
  {journal} {J. Chem. Phys.}\ }\textbf {\bibinfo {volume} {115}},\ \bibinfo
  {pages} {4450} (\bibinfo {year} {2001})}\BibitemShut {NoStop}%
\bibitem [{\citenamefont {Dattani}(2013)}]{DattaniCPC2013}%
  \BibitemOpen
  \bibfield  {author} {\bibinfo {author} {\bibfnamefont {Nikesh~S.}\
  \bibnamefont {Dattani}},\ }\bibfield  {title} {\enquote {\bibinfo {title}
  {{FeynDyn: A MATLAB} program for fast numerical {Feynman} integral
  calculations for open quantum system dynamics on {GPUs}},}\ }\href {\doibase
  https://doi.org/10.1016/j.cpc.2013.07.001} {\bibfield  {journal} {\bibinfo
  {journal} {Comput. Phys. Commun.}\ }\textbf {\bibinfo {volume} {184}},\
  \bibinfo {pages} {2828} (\bibinfo {year} {2013})}\BibitemShut {NoStop}%
\bibitem [{\citenamefont {Li}\ \emph {et~al.}(2018)\citenamefont {Li},
  \citenamefont {Hall},\ and\ \citenamefont {Wiseman}}]{LiPhR2018}%
  \BibitemOpen
  \bibfield  {author} {\bibinfo {author} {\bibfnamefont {Li}~\bibnamefont
  {Li}}, \bibinfo {author} {\bibfnamefont {Michael J.~W.}\ \bibnamefont
  {Hall}}, \ and\ \bibinfo {author} {\bibfnamefont {Howard~M.}\ \bibnamefont
  {Wiseman}},\ }\bibfield  {title} {\enquote {\bibinfo {title} {Concepts of
  quantum non-{Markovianity}: A hierarchy},}\ }\href {\doibase
  https://doi.org/10.1016/j.physrep.2018.07.001} {\bibfield  {journal}
  {\bibinfo  {journal} {Phys. Rep.}\ }\textbf {\bibinfo {volume} {759}},\
  \bibinfo {pages} {1} (\bibinfo {year} {2018})},\ \bibinfo {note} {concepts of
  quantum non-Markovianity: A hierarchy}\BibitemShut {NoStop}%
\bibitem [{\citenamefont {Milz}\ \emph
  {et~al.}(2017{\natexlab{a}})\citenamefont {Milz}, \citenamefont {Pollock},\
  and\ \citenamefont {Modi}}]{MilzOSD2017}%
  \BibitemOpen
  \bibfield  {author} {\bibinfo {author} {\bibfnamefont {Simon}\ \bibnamefont
  {Milz}}, \bibinfo {author} {\bibfnamefont {Felix~A.}\ \bibnamefont
  {Pollock}}, \ and\ \bibinfo {author} {\bibfnamefont {Kavan}\ \bibnamefont
  {Modi}},\ }\bibfield  {title} {\enquote {\bibinfo {title} {An introduction to
  operational quantum dynamics},}\ }\href {\doibase 10.1142/S1230161217400169}
  {\bibfield  {journal} {\bibinfo  {journal} {Open Sys. Info. Dyn.}\ ,\
  \bibinfo {pages} {1740016}} (\bibinfo {year}
  {2017}{\natexlab{a}})}\BibitemShut {NoStop}%
\bibitem [{\citenamefont {Nielsen}\ and\ \citenamefont
  {Chuang}(2011)}]{ChuangNielsen2011}%
  \BibitemOpen
  \bibfield  {author} {\bibinfo {author} {\bibfnamefont {Michael~A.}\
  \bibnamefont {Nielsen}}\ and\ \bibinfo {author} {\bibfnamefont {Isaac~L.}\
  \bibnamefont {Chuang}},\ }\href@noop {} {\emph {\bibinfo {title}
  {\textit{Quantum Computation and Quantum Information}}}}\ (\bibinfo
  {publisher} {Cambridge University Press},\ \bibinfo {year}
  {2011})\BibitemShut {NoStop}%
\bibitem [{\citenamefont {Wilde}(2017)}]{Wilde2017}%
  \BibitemOpen
  \bibfield  {author} {\bibinfo {author} {\bibfnamefont {Mark}\ \bibnamefont
  {Wilde}},\ }\href@noop {} {\emph {\bibinfo {title} {\textit{Quantum
  information theory; 2nd edition}}}}\ (\bibinfo  {publisher} {Cambridge
  University Press},\ \bibinfo {year} {2017})\BibitemShut {NoStop}%
\bibitem [{\citenamefont {Milz}\ \emph
  {et~al.}(2017{\natexlab{b}})\citenamefont {Milz}, \citenamefont {Sakuldee},
  \citenamefont {Pollock},\ and\ \citenamefont {Modi}}]{Milzarx2017}%
  \BibitemOpen
  \bibfield  {author} {\bibinfo {author} {\bibfnamefont {Simon}\ \bibnamefont
  {Milz}}, \bibinfo {author} {\bibfnamefont {Fattah}\ \bibnamefont {Sakuldee}},
  \bibinfo {author} {\bibfnamefont {Felix~A.}\ \bibnamefont {Pollock}}, \ and\
  \bibinfo {author} {\bibfnamefont {Kavan}\ \bibnamefont {Modi}},\ }\bibfield
  {title} {\enquote {\bibinfo {title} {Kolmogorov extension theorem for
  (quantum) causal modelling and general probabilistic theories},}\ }\href
  {https://www.arxiv.org/abs/1712.02589} {\bibfield  {journal} {\bibinfo
  {journal} {arXiv:1712.02589}\ } (\bibinfo {year}
  {2017}{\natexlab{b}})}\BibitemShut {NoStop}%
\bibitem [{\citenamefont {Pollock}\ \emph
  {et~al.}(2018{\natexlab{b}})\citenamefont {Pollock}, \citenamefont
  {Rodr\'{\i}guez-Rosario}, \citenamefont {Frauenheim}, \citenamefont
  {Paternostro},\ and\ \citenamefont {Modi}}]{Pollock2018PRL}%
  \BibitemOpen
  \bibfield  {author} {\bibinfo {author} {\bibfnamefont {Felix~A.}\
  \bibnamefont {Pollock}}, \bibinfo {author} {\bibfnamefont {C\'esar}\
  \bibnamefont {Rodr\'{\i}guez-Rosario}}, \bibinfo {author} {\bibfnamefont
  {Thomas}\ \bibnamefont {Frauenheim}}, \bibinfo {author} {\bibfnamefont
  {Mauro}\ \bibnamefont {Paternostro}}, \ and\ \bibinfo {author} {\bibfnamefont
  {Kavan}\ \bibnamefont {Modi}},\ }\bibfield  {title} {\enquote {\bibinfo
  {title} {Operational {Markov} condition for quantum processes},}\ }\href
  {\doibase 10.1103/PhysRevLett.120.040405} {\bibfield  {journal} {\bibinfo
  {journal} {Phys. Rev. Lett.}\ }\textbf {\bibinfo {volume} {120}},\ \bibinfo
  {pages} {040405} (\bibinfo {year} {2018}{\natexlab{b}})}\BibitemShut
  {NoStop}%
\bibitem [{\citenamefont {Taranto}\ \emph
  {et~al.}(2019{\natexlab{a}})\citenamefont {Taranto}, \citenamefont {Pollock},
  \citenamefont {Milz}, \citenamefont {Tomamichel},\ and\ \citenamefont
  {Modi}}]{Taranto2019PRL}%
  \BibitemOpen
  \bibfield  {author} {\bibinfo {author} {\bibfnamefont {Philip}\ \bibnamefont
  {Taranto}}, \bibinfo {author} {\bibfnamefont {Felix~A.}\ \bibnamefont
  {Pollock}}, \bibinfo {author} {\bibfnamefont {Simon}\ \bibnamefont {Milz}},
  \bibinfo {author} {\bibfnamefont {Marco}\ \bibnamefont {Tomamichel}}, \ and\
  \bibinfo {author} {\bibfnamefont {Kavan}\ \bibnamefont {Modi}},\ }\bibfield
  {title} {\enquote {\bibinfo {title} {Quantum {Markov} order},}\ }\href
  {\doibase 10.1103/PhysRevLett.122.140401} {\bibfield  {journal} {\bibinfo
  {journal} {Phys. Rev. Lett.}\ }\textbf {\bibinfo {volume} {122}},\ \bibinfo
  {pages} {140401} (\bibinfo {year} {2019}{\natexlab{a}})}\BibitemShut
  {NoStop}%
\bibitem [{\citenamefont {Taranto}\ \emph
  {et~al.}(2019{\natexlab{b}})\citenamefont {Taranto}, \citenamefont {Milz},
  \citenamefont {Pollock},\ and\ \citenamefont {Modi}}]{Taranto2019PRA}%
  \BibitemOpen
  \bibfield  {author} {\bibinfo {author} {\bibfnamefont {Philip}\ \bibnamefont
  {Taranto}}, \bibinfo {author} {\bibfnamefont {Simon}\ \bibnamefont {Milz}},
  \bibinfo {author} {\bibfnamefont {Felix~A.}\ \bibnamefont {Pollock}}, \ and\
  \bibinfo {author} {\bibfnamefont {Kavan}\ \bibnamefont {Modi}},\ }\bibfield
  {title} {\enquote {\bibinfo {title} {Structure of quantum stochastic
  processes with finite {Markov} order},}\ }\href {\doibase
  10.1103/PhysRevA.99.042108} {\bibfield  {journal} {\bibinfo  {journal} {Phys.
  Rev. A}\ }\textbf {\bibinfo {volume} {99}},\ \bibinfo {pages} {042108}
  (\bibinfo {year} {2019}{\natexlab{b}})}\BibitemShut {NoStop}%
\bibitem [{\citenamefont {Sakuldee}\ \emph {et~al.}(2018)\citenamefont
  {Sakuldee}, \citenamefont {Milz}, \citenamefont {Pollock},\ and\
  \citenamefont {Modi}}]{SakuldeeJPA2018}%
  \BibitemOpen
  \bibfield  {author} {\bibinfo {author} {\bibfnamefont {Fattah}\ \bibnamefont
  {Sakuldee}}, \bibinfo {author} {\bibfnamefont {Simon}\ \bibnamefont {Milz}},
  \bibinfo {author} {\bibfnamefont {Felix~A}\ \bibnamefont {Pollock}}, \ and\
  \bibinfo {author} {\bibfnamefont {Kavan}\ \bibnamefont {Modi}},\ }\bibfield
  {title} {\enquote {\bibinfo {title} {Non-{Markovian} quantum control as
  coherent stochastic trajectories},}\ }\href {\doibase
  10.1088/1751-8121/aabb1e} {\bibfield  {journal} {\bibinfo  {journal} {J.
  Phys. A: Math. Theor.}\ }\textbf {\bibinfo {volume} {51}},\ \bibinfo {pages}
  {414014} (\bibinfo {year} {2018})}\BibitemShut {NoStop}%
\bibitem [{\citenamefont {Atland}\ and\ \citenamefont
  {Simon}(2010)}]{AtlandSimon2010}%
  \BibitemOpen
  \bibfield  {author} {\bibinfo {author} {\bibfnamefont {A.}~\bibnamefont
  {Atland}}\ and\ \bibinfo {author} {\bibfnamefont {B.}~\bibnamefont {Simon}},\
  }\href@noop {} {\emph {\bibinfo {title} {Condensed matter field theory}}}\
  (\bibinfo  {publisher} {Cambridge University Press},\ \bibinfo {year}
  {2010})\BibitemShut {NoStop}%
\bibitem [{\citenamefont {Trotter}(1959)}]{Trotter1959PAMS}%
  \BibitemOpen
  \bibfield  {author} {\bibinfo {author} {\bibfnamefont {H.~F.}\ \bibnamefont
  {Trotter}},\ }\bibfield  {title} {\enquote {\bibinfo {title} {On the product
  of semi-groups of operators},}\ }\href
  {https://doi.org/10.1090/S0002-9939-1959-0108732-6} {\bibfield  {journal}
  {\bibinfo  {journal} {Proc. Amer. Math. Soc.}\ }\textbf {\bibinfo {volume}
  {10}},\ \bibinfo {pages} {545} (\bibinfo {year} {1959})}\BibitemShut
  {NoStop}%
\bibitem [{\citenamefont {Feynman}\ and\ \citenamefont
  {Vernon}(1963)}]{FeynmanAnP1963}%
  \BibitemOpen
  \bibfield  {author} {\bibinfo {author} {\bibfnamefont {R.~P.}\ \bibnamefont
  {Feynman}}\ and\ \bibinfo {author} {\bibfnamefont {F.~L.}\ \bibnamefont
  {Vernon}},\ }\bibfield  {title} {\enquote {\bibinfo {title} {The theory of a
  general quantum system interacting with a linear dissipative system},}\
  }\href {\doibase https://doi.org/10.1016/0003-4916(63)90068-X} {\bibfield
  {journal} {\bibinfo  {journal} {Ann. Phys.}\ }\textbf {\bibinfo {volume}
  {24}},\ \bibinfo {pages} {118} (\bibinfo {year} {1963})}\BibitemShut
  {NoStop}%
\bibitem [{\citenamefont {Di\'osi}\ and\ \citenamefont
  {Ferialdi}(2014)}]{DiosiPRL2014}%
  \BibitemOpen
  \bibfield  {author} {\bibinfo {author} {\bibfnamefont {L.}~\bibnamefont
  {Di\'osi}}\ and\ \bibinfo {author} {\bibfnamefont {L.}~\bibnamefont
  {Ferialdi}},\ }\bibfield  {title} {\enquote {\bibinfo {title} {General
  non-{Markovian} structure of {Gaussian} master and stochastic {Schr\"odinger}
  equations},}\ }\href {\doibase 10.1103/PhysRevLett.113.200403} {\bibfield
  {journal} {\bibinfo  {journal} {Phys. Rev. Lett.}\ }\textbf {\bibinfo
  {volume} {113}},\ \bibinfo {pages} {200403} (\bibinfo {year}
  {2014})}\BibitemShut {NoStop}%
\bibitem [{\citenamefont {Dattani}\ \emph {et~al.}(2012)\citenamefont
  {Dattani}, \citenamefont {Pollock},\ and\ \citenamefont
  {Wilkins}}]{DattaniQPL2012}%
  \BibitemOpen
  \bibfield  {author} {\bibinfo {author} {\bibfnamefont {Nikesh~S.}\
  \bibnamefont {Dattani}}, \bibinfo {author} {\bibfnamefont {Felix~A.}\
  \bibnamefont {Pollock}}, \ and\ \bibinfo {author} {\bibfnamefont {David~M.}\
  \bibnamefont {Wilkins}},\ }\bibfield  {title} {\enquote {\bibinfo {title}
  {Analytic influence functionals for numerical {Feynman} integrals in most
  open quantum systems},}\ }\href {\doibase http://dx.doi.org/10.18576/qpl}
  {\bibfield  {journal} {\bibinfo  {journal} {Quant. Phys. Lett.}\ }\textbf
  {\bibinfo {volume} {1}},\ \bibinfo {pages} {35} (\bibinfo {year}
  {2012})}\BibitemShut {NoStop}%
\bibitem [{\citenamefont {Schollw{\"{o}}ck}(2011)}]{Schollwock2011AP}%
  \BibitemOpen
  \bibfield  {author} {\bibinfo {author} {\bibfnamefont {Ulrich}\ \bibnamefont
  {Schollw{\"{o}}ck}},\ }\bibfield  {title} {\enquote {\bibinfo {title} {The
  density-matrix renormalization group in the age of matrix product states},}\
  }\href {\doibase https://doi.org/10.1016/j.aop.2010.09.012} {\bibfield
  {journal} {\bibinfo  {journal} {Ann. Phys.}\ }\textbf {\bibinfo {volume}
  {326}},\ \bibinfo {pages} {96} (\bibinfo {year} {2011})}\BibitemShut
  {NoStop}%
\bibitem [{Note1()}]{Note1}%
  \BibitemOpen
  \bibinfo {note} {This correspondence is not precise, however, and our usage
  of `(non-)local' should not be confused with that in the context of memory
  kernel convolution.}\BibitemShut {Stop}%
\bibitem [{\citenamefont {Yuen-Zhou}\ \emph {et~al.}(2014)\citenamefont
  {Yuen-Zhou}, \citenamefont {Krich}, \citenamefont {Kassal}, \citenamefont
  {Johnson},\ and\ \citenamefont {Aspuru-Guzik}}]{KassalSpectroscopy2014}%
  \BibitemOpen
  \bibfield  {author} {\bibinfo {author} {\bibfnamefont {Joel}\ \bibnamefont
  {Yuen-Zhou}}, \bibinfo {author} {\bibfnamefont {Jacob~J}\ \bibnamefont
  {Krich}}, \bibinfo {author} {\bibfnamefont {Ivan}\ \bibnamefont {Kassal}},
  \bibinfo {author} {\bibfnamefont {Allan~S}\ \bibnamefont {Johnson}}, \ and\
  \bibinfo {author} {\bibfnamefont {Al\'{a}n}\ \bibnamefont {Aspuru-Guzik}},\
  }\href {\doibase 10.1088/978-0-750-31062-8} {\emph {\bibinfo {title}
  {Ultrafast Spectroscopy}}}\ (\bibinfo  {publisher} {IOP},\ \bibinfo {year}
  {2014})\BibitemShut {NoStop}%
\bibitem [{\citenamefont {Abramowitz}\ and\ \citenamefont
  {Stegun}(1964)}]{Abramowitz1964}%
  \BibitemOpen
  \bibfield  {author} {\bibinfo {author} {\bibfnamefont {Milton}\ \bibnamefont
  {Abramowitz}}\ and\ \bibinfo {author} {\bibfnamefont {Irene~A.}\ \bibnamefont
  {Stegun}},\ }\href@noop {} {\emph {\bibinfo {title} {Handbook of Mathematical
  Functions}}}\ (\bibinfo  {publisher} {New York: Dover Publications},\
  \bibinfo {year} {1964})\BibitemShut {NoStop}%
\end{thebibliography}%

\newpage
\appendix

\section{Process tensor formalism.}
\label{app:proc}

As described in the main text, we consider the scenario characterized by Eq.~\eqref{eq:QuantumProcess}, where an open quantum system $S$ is periodically interrogated as it evolves. The slightly more general case involves an arbitrary time separation between interventions and a possibly time-dependent $SE$ Hamiltonian $H(t)$, such that the state at the $k^{\rm th}$ time is given by
\begin{gather} \label{eq:generalprocess}
    \rho_k(\{\mathcal{A}_j\}_{j=0}^{k-1}) = \tr_{E} \left\lbrace \mathcal{U}_{k:k-1}\mathcal{A}_{k-1} \ ...
        \ \mathcal{U}_{1:0} \mathcal{A}_{0} \left[ \chi_{0} \right] \right\rbrace \,
\end{gather}
where $\mathcal{U}_{j:j-1}$ is the time-evolution superoperator from time $t_{j-1}$ to $t_j$ with action $\mathcal{U}_{j:j-1}[\rho] = U_{j:j-1} \rho U_{j:j-1}^\dagger$ in terms of the time-ordered exponential 
\begin{gather} 
    U_{j:j-1} = T_{\leftarrow}\exp\left\{-i\int_{t_{j-1}}^{t_j} ds\, H(s) \right\}.
\end{gather} 
As in the main text, $\chi_0$ is the potentially correlated initial $SE$ state and each $\mathcal{A}_j$ can be any superoperator, though only those that are completely positive correspond to physically realizable transformations~\cite{MilzOSD2017}. 

Any superoperator can be represented in terms of an operator sum as $\mathcal{A}[\rho] = \sum_{n} X_n \rho Y_n^\dagger$ (completely positive maps are characterized by $X_n=Y_n$, in which case the latter are called Kraus operators). As such, by taking the trace over the final state in Eq.~\eqref{eq:generalprocess}, any multi-time correlation function $\langle B_0(t_0) \dots B_{k-1}(t_{k-1}) A_{k-1}(t_{k-1}) \dots A_0(t_0)\rangle_{\chi_0}$, where $\{A_j(t_j) = U_{j:0}^\dagger A_j U_{j:0}\}$ and $\{B_j = U_{j:0}^\dagger B_j U_{j:0}\}$ are Heisenberg picture operators on $S$, can be obtained by choosing $\mathcal{A}_j[\rho] = A_j \rho B_j $. In this way, with a sufficiently dense set of times $\{t_j\}$, any dynamically observable property of $S$ can be represented in the form of Eq.~\eqref{eq:generalprocess} (correlation functions involving fewer observables can be obtained by choosing some of the $\{A_j\}$ and $\{B_j\}$ to be the identity operator). In addition, the freely evolved final state $\rho_k$ can be obtained by choosing all $\mathcal{A}_j=\mathcal{I}$, where the latter is the identity superoperator with action $\mathcal{I}[\rho] = \rho$.

As we will now show, all this information can be encoded in a single object, the process tensor. While it is often introduced in terms of an abstract multi-linear map, we will here express it solely in terms of a concrete matrix representation via a version of the Choi-Jamio{\l}kowski isomorphism. The Choi state (or Choi matrix) $\mathsf{A}$ of a superoperator $\mathcal{A}$ is defined in terms of its action on one half of the (unnormalized) maximally entangled state $\Psi = \sum_{sr}\ketbra{ss}{rr}$, where $\{\ket{s}\}$ forms an orthonormal basis for the $d$-dimensional system, as
\begin{equation}
    \mathsf{A} = \mathcal{A} \otimes \mathcal{I} [\Psi] = \sum_{srs'r'} \mathcal{A}^{(s,r,s',r')}\ketbra{ss'}{rr'},
\end{equation}
with $\mathcal{A}^{(s,r,s',r')} = \bra{s}\mathcal{A}[\ketbra{s'}{r'}]\ket{r}$.
Labelling the first and second copies of the original system's Hilbert space $\mathtt{o}$ (for output) and $\mathtt{i}$ (for input) respectively, the action of the superoperator on an initial state $\rho$ can be written in terms of this representation as $\mathcal{A}[\rho] = \tr_\mathtt{i}[\mathsf{A} \,\mathbbm{1}_\mathsf{o}\otimes \rho^T ]$, where the trace is over the input subsystem.

By expanding out the action of the superoperators in Eq.~\eqref{eq:generalprocess} and inserting a resolution of the identity on $S$ to the left and right of every unitary matrix, one arrives at the equivalent expression
\begin{gather}
    \rho_k(\{\mathcal{A}_j\}_{j=0}^{k-1}) = \tr_{k-1:0}\left\{\Upsilon_{k:0}(\mathbbm{1}_k\otimes \mathbf{A}^T_{k-1:0})\right\},
\end{gather}
with $\mathbf{A}_{k-1:0} = \mathsf{A}_{k-1}\otimes \cdots \otimes \mathsf{A}_1\otimes\mathsf{A}_0$ and the trace over all (input and output) Hilbert spaces on which $\mathbf{A}_{k-1:0}$ acts. The positive operator $\Upsilon_{k:0}$ is (the Choi state of) the process tensor, and it can be expressed in terms of the underlying $SE$ dynamics as
\begin{align} \label{eq:generalptchoi}
    \Upsilon_{k:0}  =&\!\!\!\!   \sum_{\vec{s}',\vec{r}',\vec{s},\vec{r}}\!\!\!\tr \left\lbrace  \mathcal{U}_{k:k-1}^{(s'_{k}, r'_{k},s_{k-1}, r_{k-1})} \!\!\!\!\!\! \dots \ \mathcal{U}_{1:0}^{(s'_1, r'_1,s_0, r_0)} \!\left[ \chi^{(r'_0,s'_0)}_0 \right]  \right\rbrace \nonumber\\
	    & \qquad\!\!\times \ketbra{s'_k s_{k-1}\dots s'_1 s_0 s'_0}{r'_k r_{k-1}  \dots r'_1 r_0 r'_0},\!\!
\end{align}
with $\mathcal{U}_{j:j-1}^{(s', r',s, r)}[\rho^E] = \bra{s'}U_{j:j-1} (\ketbra{s}{r} \otimes \rho^E) U^\dagger_{j:j-1} \ket{r'}$ and  $\chi^{(r',s')}_0 = \bra{r'}\chi_0\ket{s'}$. This object can be directly constructed by swapping the system with one half of a maximally entangled state at each point where a superoperator $\mathcal{A}$ is to be applied~\cite{Pollock2018PRA}. When the $SE$ Hamiltonian is time independent and $t_j-t_{j-1} = \delta t$ for all $j$, Eq.~\eqref{eq:generalptchoi} is equivalent to Eq.~\eqref{eq:processtensor} of the main text.

A representation in terms of the process tensor separates the process and external interventions, as illustrated in Fig.~\ref{fig:StateFunction} of the main text. The process tensor's properties reflect the necessary features of any physical open dynamics: $\Upsilon_{k:0}$ is positive if and only if the process is completely positive, and causality is encoded in the hierarchy of trace conditions $\tr_{j}\Upsilon_{j:0} = \Upsilon_{j-1:0}\otimes \mathbbm{1}_{\mathtt{o}_{j-1}}$.
While we have expressed it in terms of $SE$ quantities, it is an operator only on copies of the Hilbert space of $S$ and it  has a natural matrix product form, allowing for an efficient representation in many cases, a fact we exploit in this paper.

\section{Connection with influence functional}
\label{app:inffunc}

In the case where the time spacing is small (and the Hamiltonian varies relatively slowly), the Trotter formula can be used to approximate the time evolution superoperators as $\mathcal{U}_{j:j-1} \simeq \mathcal{V}^{1/2}_{j:j-1} \mathcal{W}_{j:j-1}\mathcal{V}^{1/2}_{j:j-1}$, with $\mathcal{V}_{j:j-1}$ generated by the $S$ part of the Hamiltonian and $\mathcal{W}_{j:j-1}$ by the remainder. Expanding out the superoperators appearing inside the trace in Eq.~\eqref{eq:generalptchoi} and introducing further resolutions of the identity, one finds
\begin{align} \label{eq:trotteru}
    \mathcal{U}_{j:j-1}^{(s',r',s,r)} \simeq& \bra{s'}\mathcal{V}^{1/2}_{j:j-1} \mathcal{W}_{j:j-1}\mathcal{V}^{1/2}_{j:j-1}\left[\ketbra{s}{r}\right] \ket{r'}\nonumber \\
    =& \sum_{t',t,u',u}\bra{s'}\mathcal{V}^{1/2}_{j:j-1}\left[\ketbra{t'}{u'}\right] \ket{r'} \nonumber \\
    & \qquad\times\bra{s}{\mathcal{V}^{*\,1/2}_{j:j-1}}\left[\ketbra{t}{u}\right] \ket{r} \mathcal{W}^{(t',u',t,u)}_{j:j-1},
\end{align}
where $\mathcal{W}^{(s',r',s,r)}_{j:j-1}:= \bra{s'}\mathcal{W}_{j:j-1}\left[\ketbra{s}{r}\right]\ket{r'}$, and we have used that $\bra{t}\mathcal{V}^{1/2}_{j:j-1}\left[\ketbra{s}{r}\right] \ket{u} = \bra{s}{\mathcal{V}^{*\,1/2}_{j:j-1}}\left[\ketbra{t}{u}\right] \ket{r}$. Assuming a factorizing initial condition $\chi_0=\rho_0 \otimes \tau$ and substituting Eq.~\eqref{eq:trotteru} into Eq.~\eqref{eq:generalptchoi} leads to the following slightly more general version of Eq.~\eqref{eq:influencefunc} for the approximate process tensor $\tilde{\Upsilon}_{k:0}$:
\begin{gather}
    \tilde{\Upsilon}_{k:0} = \bigotimes_{j=1}^k\left(\mathcal{V}^{1/2}_{j:j-1}\otimes\mathcal{V}^{*\,1/2}_{j:j-1}\right)\left[\mathcal{F}_{k:0}\right] \otimes \rho_0,
\end{gather}
with
\begin{align} \label{eq:generalinf}
    \mathcal{F}_{k:0} =&\!\! \sum_{\vec{s}',\vec{r}',\vec{s},\vec{r}}\!\!\tr_{E} \left\lbrace  \mathcal{W}_{k:k-1}^{(s_{k}', r_{k}',s_{k-1}, r_{k-1})}  \!\dots \mathcal{W}_{1:0}^{(s_1', r_1',s_0, r_0)} \left[ \tau \right]  \right\rbrace\nonumber\\
	    & \quad\times \ketbra{s_k' s_{k-1} \dots s_1 s_1' s_0}{r_k' r_{k-1} \dots r_1 r_1' r_0}.
\end{align}

In the special case that the bath coupling part Hamiltonian can be written in the form $H_B(t) = \sum_s \ketbra{s}{s} \otimes B_s(t)$ where $\sum_s \ketbra{s}{s} = \mathbbm{1}_S$ (the Hamiltonian of the main text takes this form in the interaction picture with respect to the bath), $\mathcal{W}_{j:j-1}^{(s', r',s, r)} = \delta_{ss'} \delta_{rr'} \mathcal{W}_{j:j-1}^{(s, r)}$ with 
\begin{gather} \label{eq:indexedsuper}
    \mathcal{W}_{j:j-1}^{(s, r)}[\rho^E] = W_{j:j-1}^{(s)} \rho^E W_{j:j-1}^{(r)\, \dagger}
\end{gather} 
and
\begin{gather}
    W_{j:j-1}^{(s)}=T_{\leftarrow}\exp[-i\int_{t_{j-1}}^{t_j} dx\, B_s(x)].
\end{gather}
For a time-independent Hamiltonian with even time spacing $\delta t$, Eq.~\eqref{eq:generalinf} then reduces to the operator representation of the discretized Feynman-Vernon influence functional in Eq.~\eqref{eq:infuncstate} of the main text.

\section{Explicit form of influence tensors in the spin-boson model}
\label{app:spinboson}

Further decomposing the influence functional into a product of the form of Eq.~\eqref{eq:influencefunc} requires that the bath be composed of field modes coupled linearly to the system, and that the Hamiltonian and initial state are quadratic in the corresponding creation and annihilation operators $\{\hat{a}^\dagger_n\}$ and $\{\hat{a}_n\}$. In other words, the environment must be Gaussian and the operators coupling to the system must take the form $\hat{B}_s(t) = \sum_n (g_{s,n} \hat{a}_n e^{- i \omega_n t} + g_{s,n}^* \hat{a}_n^\dagger e^{i \omega_n t})$ in the interaction picture. In this case, Wick's theorem can be applied to express Eq.~\eqref{eq:generalinf} as a product of exponentiated two point correlation functions. Specifically, we use the fact that for any linear functional of bath operators $\hat{X}$, $ \tr\{T\exp[\hat{X}] \rho\} = \exp[\frac{1}{2}\tr\{T\hat{X}^2 \rho\}]$, with $T$ any time ordering operator, when $\rho$ is Gaussian~\cite{DiosiPRL2014}. Treating the left and right appended operators in Eq.~\eqref{eq:indexedsuper} as a single contour ordered exponential under the trace, this results in the following expression:
\begin{align}
    \mathcal{F}_{k:0}^{\alpha_{k}...\alpha_{1}}=&\tr_{E} \left\lbrace  \mathcal{W}_{k:k-1}^{(s_{k}, r_{k})}  \!\dots \mathcal{W}_{1:0}^{(s_1, r_1)} \left[ \tau \right]  \right\rbrace \nonumber\\
    =& \exp\left[-\frac{1}{2}\sum_{i\geq j} \left(\zeta^{(s_i, s_j)}_{i,j} + \zeta^{(r_i, r_j)\, *}_{i,j} \right. \right. \nonumber \\
    & \qquad\qquad\qquad\left. \vphantom{-\frac{1}{2}\sum_{i\geq j}} \left.- \zeta^{(r_i, s_j)}_{i,j} - \zeta^{(s_i, r_j)\,*}_{i,j}\right)\right], \label{eq:gaussianinf}
\end{align}
where
\begin{gather}
    \zeta^{(u, v)}_{i,j} = \int_{t_{i-1}}^{t_i}dx\int_{t_{j-1}}^{t_j} dy \tr\left\{\hat{B}_u(x)\hat{B}_v(y) \tau\right\} \label{eq:zeta}
\end{gather}
for $i\neq j$, else for $i=j$:
\begin{gather}
    \zeta^{(u, v)}_{i,i} = \int_{t_{i-1}}^{t_i}dx\int_{t_{i-1}}^{x} dy \tr\left\{\hat{B}_u(x)\hat{B}_v(y) \tau\right\}.  \label{eq:xi}
\end{gather}

Restricting to the spin-boson type Hamiltonian considered in the main text, with a single interaction term $\hat{s}\sum_n(g_n \hat{a}_n + g_n^* \hat{a}_n^\dagger)$ and with a thermal (and hence Gaussian) initial bath state $\tau_\beta$, we can compute the influence tensors explicitly. Here, the interaction picture bath operators appearing in Eqs.~\eqref{eq:zeta}~and~\eqref{eq:xi} take the simple form $\hat{B}_u(t) = \lambda_u \sum_n (g_{n} \hat{a}_n e^{- i \omega_n t} + g_{n}^* \hat{a}_n^\dagger e^{i \omega_n t})$, written in terms of the eigenvalues of the system operator $\hat{s} = \sum_u \lambda_u \ketbra{u}{u}$.

In this case, $\zeta^{(u, v)}_{i,j} = 2\lambda_u \lambda_v \eta_{i-j}$, where the memory kernel elements 
\begin{equation}
	\eta_{i-j} = 
	\begin{cases}
	\int_{t_{i-1}}^{t_{i}} \int_{t_{j-1}}^{t_{j}} dt' dt'' C(t'-t'') \ ,& \ \ i \neq j \\
	\int_{t_{i-1}}^{t_{i}} \int_{t_{i-1}}^{t'} dt' dt'' C(t'-t'') \ ,& \ \ i = j
	\end{cases},
\end{equation}
depend only on the difference between time steps and are expressed in terms of the environment auto-correlation function~\cite{Strathearn2018NC,Strathearn2017NJP}
\begin{align} \label{eq:bathcorrfunc}
    C(t) = \frac{1}{\pi} \int_{0}^{\infty} d \omega J(\omega) 
    \frac{\cosh \left[ \omega \left( \beta/2-it \right) \right]}{ \sinh \left[ \beta \omega /2 \right]}.
\end{align}
Here $J(\omega) = \sum_{n}|g_n|^2 \delta(\omega - \omega_n)$ is the spectral density, as defined in the main text. In terms of these quantities, the elements of the influence tensors $b_{(i-j)}$ that correspond to each of the terms in the exponentiated sum in Eq.~\eqref{eq:gaussianinf} (such that $\mathcal{F}_{k:0}^{\alpha_{k}...\alpha_{1}} = \prod_{i\geq j} [b_{(i-j)}]^{\alpha_i \alpha_j}$), can be written 
\begin{equation} \label{eq:inftensors}
	\left[ b_{(i-j)} \right]^{\alpha_{i} \alpha_{j}} = e^{ -(\lambda_{s_{i}}-\lambda_{r_{i}})( \eta_{i-j}\lambda_{s_{j}}-\eta_{i-j}^{*}\lambda_{r_{j}} )}.
\end{equation}
These are the values that enter directly into our algorithm.

\section{Tensor network compression} 
\label{app:algorithm}

\begin{figure*}
	\includegraphics[width=17cm]{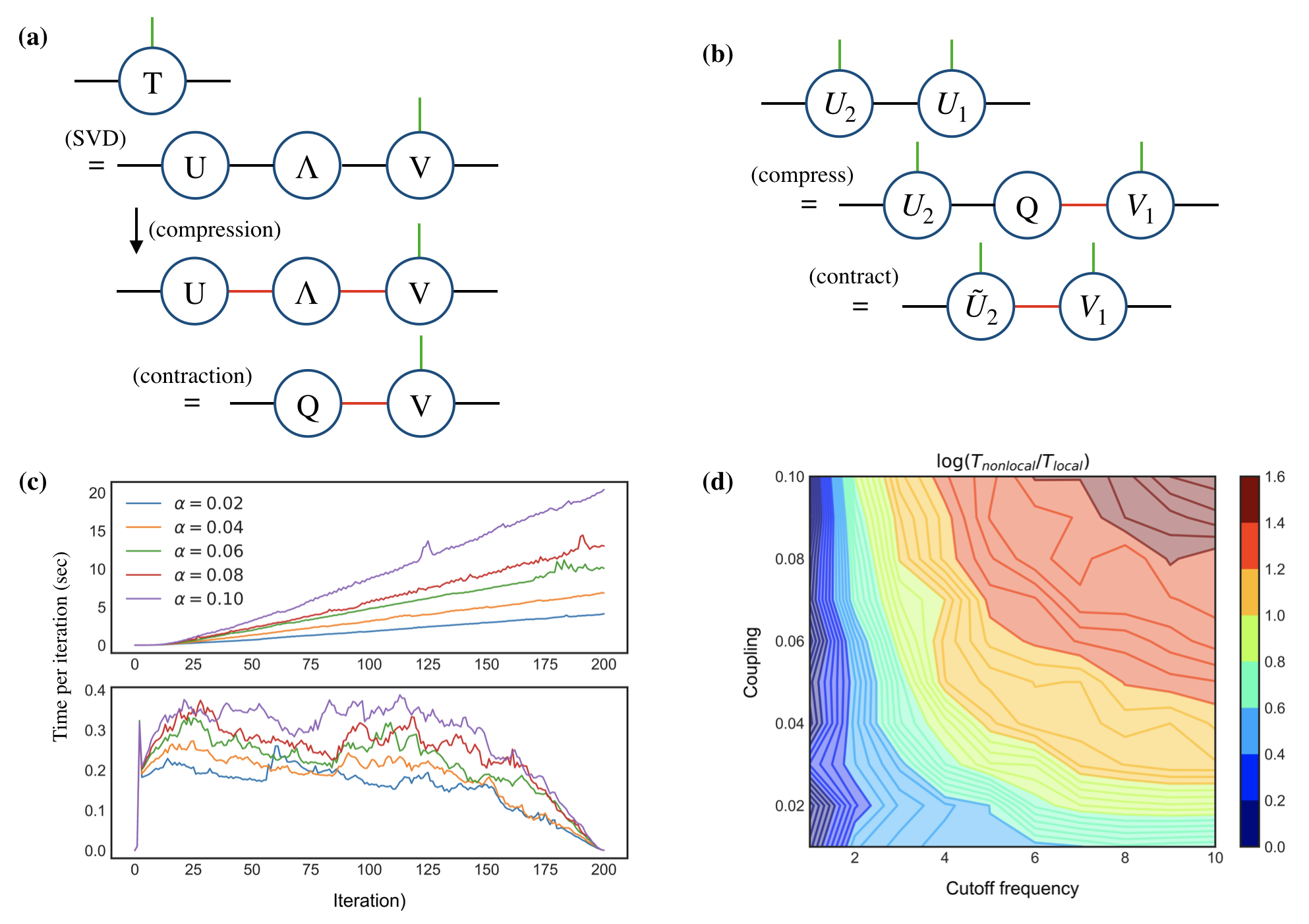}
	\caption{(a) Singular value decomposition applied to input tensor.
	The singular values are truncated, giving a smaller bond dimension (indicated by red color).
	A new local tensor is defined and the contribution $q_{c}$ is ready to be propagated along the matrix product state.
	(b) A local tensor of the boundary MPS is subjected to a SVD compression.
	The compression procedure is repeated iteratively across the full MPS.
	Here we illustrate a left compression sweep, the full procedure includes compression sweeps in both directions.
	(c) Comparison between the computation times per iteration for the local and non-local algorithms at weak coupling.
    The parameters used are $\omega_{c} = 5 \Omega$, $\nu=1$, $\delta t = 0.04/\Omega$ and a singular value cutoff of
    $\lambda_{c} = 10^{-6}$.
    We see that the non-local algorithm has a computation time increasing linearly with each iteration,
    this is contrasted with the local algorithm where an approximately constant computation time is observed.
    In addition we find a significant improvement in the actual value of the computation time,
    as argued in the main text this is due to a separation between relevant and irrelevant information in the compression.
    (d) Log ratio of total computation time for the non-local ($T_{nonlocal}$) and the local ($T_{local}$) algorithms for a range of couplings $\alpha$ and cutoff frequencies $\omega_c$. These calculations were performed at zero temperature and otherwise for the same parameters as in Fig.~\ref{fig:complexity}.}
	\label{fig:compression}
\end{figure*}

In contracting the network, efficiency is achieved by finding a minimal approximate representation
for the boundary matrix product operator in each iteration~\cite{Schollwock2011AP}.
This is obtained by replacing high rank tensors with small singular values by lower rank approximations.
These are found by performing a singular value decomposition (SVD) of the local tensors in the matrix product state,
and discarding the singular values below a cutoff $\lambda_c$.
Indicating an index partition by raised and lowered indices, the SVD decomposed tensor takes the form
\begin{equation}
    \mathcal{F}^{\alpha_{k}...\alpha_{j+1}}{}_{\alpha_{j}...\alpha_{i}} = U^{\alpha_{k}...\alpha_{j+1}}{}_{\gamma} 
    \ \Lambda^{\gamma} {}_{\delta} \ \left( V^{\dagger}\right){}^{\delta} {}_{\alpha_{j}...\alpha_{i}} \ ,
\end{equation}
where the diagonal matrix $\Lambda^{\gamma}{}_{\delta}$ contains the singular values, and $U$, $V$ are rectangular isometric matrices satisfying $U^\dagger U = \mathbbm{1}$ and $V^\dagger V = \mathbbm{1}$ (see Fig. \ref{fig:compression}a).
Truncating the singular values reduces the sizes of $U$ and $V$ (we will refer to the truncated versions as $\bar{U}$ and $\bar{V}$), and introduces a corresponding truncation error,
whose magnitude is determined by the cutoff.
We truncate the singular values such that
\begin{equation}
    \lambda_{c} \leq \sqrt{\frac{\Lambda^{2} - \tilde{\Lambda}^{2}}{\Lambda^{2}}} ,
\end{equation}
where $\tilde{\Lambda}$ denotes the truncated diagonal matrix.
Having truncated the singular values,
we contract the $\bar{U}$ and $\tilde{\Lambda}$ to give a new matrix $Q$ with which to express the newly compressed local tensor:
\begin{equation}
    \mathcal{F}^{\alpha_{k}...\alpha_{j+1}}{}_{\alpha_{j}...\alpha_{i}} \simeq 
    Q^{\alpha_{k}...\alpha_{j+1}}{}_{\delta} \  \left( \tilde{V}^{\dagger}\right){}^{\delta} {}_{\alpha_{j}...\alpha_{i}} .
\end{equation}

The singular value compression proceeds from one end of the boundary matrix product operator.
Say we begin the compression at the right boundary (see Fig.~\ref{fig:network}), then
initially the first (farthest to the right) local tensor is singular value decomposed and compressed.
The matrix $\bar{U}$ in the above decomposition is then contracted with the diagonal matrix $\bar{\Lambda}$ to give $Q$, which is subsequently contracted with the second local tensor to the left.
The compressed $\bar{V}$ tensor is stored as the new first local tensor (see Fig.\ref{fig:compression}b) and this procedure is repeated for the second local tensor, and so on until the left boundary is reached.
This constitutes a left sweep of the SVD compression procedure.
After this left sweep, an equivalent right sweep is performed, where the left-most tensor is singular value decomposed and compressed, followed by the next left-most and so on.
To produce the figures in this paper, we implemented one left sweep and one right sweep in each stage of the algorithm.
Including more sweeps back and forth would in principle improve the quality of the compression;
however, as we discuss below our implementation is sufficient to demonstrate an improvement of the local over the non-local algorithm.

\section{Scaling of memory effects with physical parameters}
\label{app:scaling}

We now proceed to estimate the complexity of contracting the network. This depends crucially on how quickly the memory decays, and hence the effective depth of the tensor network in Fig.~\ref{fig:network}c (as we will see, the overall size of memory effects is also important). From Eq.~\eqref{eq:inftensors}, it is clear that non-trivial contributions of the $b_{(i-j)}$ tensors to the influence functional depend on the magnitude of the memory kernel elements $\eta_{i-j}$. 
Specifically, the effect of truncating the network in Fig.~\ref{fig:network}c at a depth $m$, with resulting influence functional elements $[\mathcal{F}^{\alpha_k\dots \alpha_1}_{k:0}]_m$, is to introduce a relative error: 
\begin{align} \label{eq:errorbound}
    \varepsilon_m := & \frac{\mathcal{F}^{\alpha_k\dots \alpha_1}_{k:0}}{[\mathcal{F}^{\alpha_k\dots \alpha_1}_{k:0}]_m}-1 = \prod_{i=1}^{k-m}\prod_{j=1}^{i}\left[b_{(i-j+m)}\right]^{\alpha_{i+m}\alpha_j}-1 \nonumber \\
    \simeq & \sum_{i=1}^{k-m}\sum_{j=1}^{i}(\lambda_{r_{i+m}}-\lambda_{s_{i+m}})(\eta_{i-j+m} \lambda_{s_j} - \eta^*_{i-j+m} \lambda_{r_j}) \nonumber \\
    \leq & 
    2\|\hat{s}\|_{\rm op} \sum_{l=m}^{k}(k-l)|\eta_{l}|,
\end{align}
where $\|\hat{s}\|_{\rm op}:= \max{\{|\lambda_r|\}}$ (we will henceforth take $\|\hat{s}\|_{\rm op} =1$, effectively absorbing it into the coupling strength); in the second line we have assumed $m$ is sufficiently large that the error is small. 
For a fixed error $\varepsilon$, the memory time $t_m = m \delta t$, and hence the complexity of our algorithm (we expect the error due to SVD compression to scale similarly), will therefore depend on how quickly $|\eta_{l}|$ decays with $l$. If it decays exponentially with rate $c$, then in the limit of large $k$, it is relatively straightforward to show that the memory time scales as $t_m\sim \delta t (\log k + \log \varepsilon^{-1})/c$. However, as we will now see, for the spectral density we have chosen, the memory kernel decays as a power law. 

When $\delta t = t_j - t_{j-1}$ is sufficiently small, we have
\begin{align}\label{eq:smalldt}
    \eta_{i-j} \simeq \begin{cases}
    \delta t^2 C\left((i-j)\delta t\right) \ ,& \ \ i \neq j \\
	\frac{1}{2}\delta t^2 
	C(0) \ ,& \ \ i = j
	\end{cases}.
\end{align}
For the spectral density introduced in the main text $J(\omega) = (\alpha \omega_c/2) (\omega/\omega_{c})^{\nu} \exp \left( - \omega/\omega_{c} \right)$ with $\nu=1$ (i.e. the Ohmic case), the integral in Eq.~\eqref{eq:bathcorrfunc} can be evaluated explicitly, giving
\begin{align} \label{eq:analytic}
     C(t) = \frac{\alpha \omega_c^2}{2 \pi}&\left( \frac{\omega_c^2 t^2-1}{(\omega_c^2 t^2+1)^2} + \frac{2}{\beta^2\omega_c^2}\Re\psi^{(1)}\left[\frac{1-i\omega_c t}{\beta \omega_c}\right]\right. \nonumber \\ &\;\;\left.  -2 i \frac{\omega_c t}{(\omega_c^2 t^2+1)^2}\right),
\end{align}
with $\psi^{(1)}[z] := \int_{0}^\infty\, d x\, x e^{-z x}/(1-e^{-x})$ the order-1 polygamma function. For large $|z|$, the latter goes as $\psi^{(1)}[z]\sim 1/z + 1/(2z^2)$~\cite{Abramowitz1964}. 
Hence, in the limit that $ t\gg \omega_c^{-1}$ and $t \gg \beta$, we can expand out Eq.~\eqref{eq:analytic} and combine with Eq.~\eqref{eq:smalldt} to arrive at
\begin{gather}
    |\eta_{i-j}| = \frac{\alpha}{\pi\beta \omega_c |i-j|^2} + \mathcal{O}(|i-j|^{-4}),
\end{gather}
to leading order in $|i-j|^{-1}$. Therefore, for sufficiently large $m$ and $k = t_{\rm max}/\delta t$, we can perform the sum in Eq.~\eqref{eq:errorbound}, finding $\varepsilon_m \lesssim \alpha k \psi^{(1)}[m]/(\pi \beta \omega_c) \simeq \alpha k/(\pi \beta \omega_c m)$. For fixed error $\varepsilon$, we therefore have that the bound on the memory time, and hence the complexity of the algorithm scales as
\begin{gather}
    t_m \sim \frac{\alpha t_{\rm max}}{\pi \beta \omega_c \varepsilon}.
\end{gather}

This explains the behaviour of Fig.~\ref{fig:complexity}a and the large $\omega_c$ behaviour of Fig.~\ref{fig:complexity}b in the main text.
However, the onset of this limit depends on the parameter combinations $\omega_c t$ and $\sqrt{(t^2+\omega_c^{-2})/\beta^2}$ both being large (these elicit expansions for the first and second terms of Eq.~\eqref{eq:analytic} respectively). When $\omega_c \lesssim t_{\rm max}^{-1}$, the former limit is never reached, and 
only the term involving the polygamma function contributes significantly to the error. Specifically, the magnitude of the memory kernel is approximately constant, going as
\begin{gather}
    |\eta_{i-j}| = \frac{\alpha \omega_c \delta t^2 }{\pi\beta} 
    + \mathcal{O}(\omega_c^2).
\end{gather}
Therefore, for small $\delta t$, Eq.~\eqref{eq:errorbound} leads to $\varepsilon_m \lesssim (\alpha \omega_c /\pi \beta) (t_{\rm max} - t_m)^2$ and one can see that, as long as $t_{\rm max}$ is fixed, the error is bounded by a number that goes to zero as $\omega_c$ does, even for very small memory times.
Hence, in this limit, the effective coupling to the bath is weak overall and even the local influence tensors do not contribute significantly to the dynamics, explaining the behaviour at small $\omega_c$ in Fig.~\ref{fig:complexity}b.

\section{Scaling comparison}
\label{app:efficiency}

In Fig.~\ref{fig:compression}c we compare the computation time per iteration of the non-local and the local network representations
for a fixed network size and SVD cutoff.
We see that the time per iteration of the non-local TEMPO algorithm increases approximately linearly.
For a finite network, the local TEMPO algorithm has a time per iteration which rapidly goes to a non-increasing value.
The growth of complexity in the non-local case is mainly due to the build up of irrelevant information,
rather than a genuine build-up of temporal correlations.
The observed decrease in the computation time per iteration for the local case,
is a consequence of working with a finite network.
In the original TEMPO proposal \cite{Strathearn2018NC}, it was argued that one could implement a truncation of the number of tensors in the propagators and obtain a constant scaling at long times.
The same method could be applied with the local TEMPO algorithm, only with a significantly reduced time per iteration.
More rigorously, we could combine the tools developed here with the transfer tensor approach \cite{Cerrillo2014PRL,Pollock2018Qtm},
which infers long-time correlations from a short-time simulation.
The problem would then become efficiently contracting the full network up to a sufficiently long-time.

The advantage in contracting the network persists across a wide range of parameters, as depicted in Fig.~\ref{fig:compression}d. Even in the easier regime of weak coupling and small cutoff frequency, the non-local network takes longer to contract than the local one.

All the simulations presented in this work were carried out on a
2011 MacBook Pro with a 2,4 GHz Intel Core i5 processor and a 4GB 1333 MHz DDR3 memory.
The algorithm has been implemented using the programming language Python,
no specialized packages, beyond NumPy, were used.
There is extensive scope for optimization, and utilizing packages for tensor network manipulations would supposedly make the implementation quite simple.

\end{document}